\documentclass[acmtog, nonacm]{acmart}
\acmSubmissionID{0271}

\usepackage{cirl}
\usepackage{volumetric-wos}

\setcitestyle{nosort,square} 

\title{Solving partial differential equations in participating media}

\author{Bailey Miller}
\email{bmmiller@andrew.cmu.edu}
\affiliation{%
  \institution{Carnegie Mellon University}
  \streetaddress{5000 Forbes Ave}
  \city{Pittsburgh}
  \state{PA}
  \postcode{15213}
  \country{USA}
}
\orcid{0009-0009-0881-0351}

\author{Rohan Sawhney}
\email{rsawhney@nvidia.com}
\affiliation{
  \institution{NVIDIA}
  \streetaddress{2788 San Tomas Expy}
  \city{Santa Clara}
  \state{CA}
  \postcode{95051}
  \country{USA}
}
\orcid{0000-0002-3661-1554}

\author{Keenan Crane}
\email{kmcrane@cs.cmu.edu}
\affiliation{
  \institution{Carnegie Mellon University}
  \streetaddress{5000 Forbes Ave}
  \city{Pittsburgh}
  \state{PA}
  \postcode{15213}
  \country{USA}
}
\orcid{0000-0003-2772-7034}

\author{Ioannis Gkioulekas}
\email{igkioule@cs.cmu.edu}
\affiliation{%
  \institution{Carnegie Mellon University}
  \streetaddress{5000 Forbes Ave}
  \city{Pittsburgh}
  \state{PA}
  \postcode{15213}
  \country{USA}
}
\orcid{0000-0001-6932-4642}

\renewcommand\shortauthors{Miller, Sawhney, Crane, Gkioulekas}

\acmJournal{TOG}
\acmYear{2025} \acmVolume{44} \acmNumber{4} \acmArticle{} \acmMonth{8} \acmPrice{}\acmDOI{10.1145/3731152}

\begin{teaserfigure}
	\centering
	\includegraphics{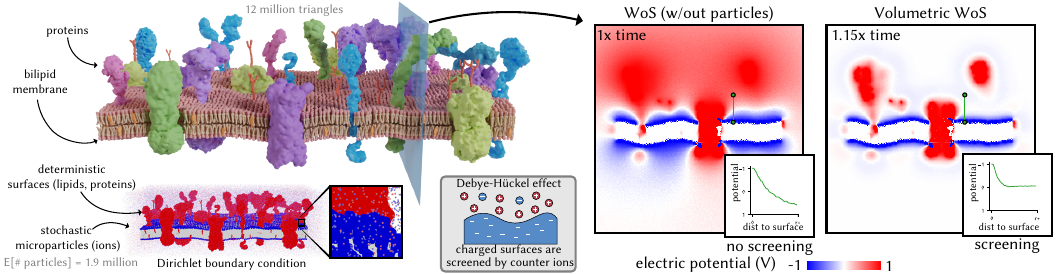}
	\caption{Real physical systems, such as biomembranes, have extraordinarily complex geometry, often managed by \emph{homogenizing} particulate substances that mediate physical interactions. For example, in the setup we show, a homogenized PDE is commonly used to model \emph{electrostatic screening} due to ions in solution. In reality, however, homogenization can be highly inaccurate, as particles often exhibit a scale similar to geometric features. We develop Monte Carlo methods that directly account for both particle and boundary geometry, maintaining efficiency without any limiting assumptions or geometric simplification.\label{fig:Membrane}}
\end{teaserfigure}

\begin{abstract}
	We consider the problem of solving partial differential equations (PDEs) in domains with complex microparticle geometry that is impractical, or intractable, to model explicitly. Drawing inspiration from volume rendering, we propose tackling this problem by treating the domain as a participating medium that models microparticle geometry \emph{stochastically}, through aggregate statistical properties (e.g., particle density). We first introduce the problem setting of PDE simulation in participating media. We then specialize to \emph{exponential media} and describe the properties that make them an attractive model of microparticle geometry for PDE simulation problems. We use these properties to develop two new algorithms, \emph{volumetric walk on spheres} and \emph{volumetric walk on stars}, that generalize previous Monte Carlo algorithms to enable efficient and discretization-free simulation of linear elliptic PDEs (e.g., Laplace) in participating media. We demonstrate experimentally that our algorithms can solve Laplace boundary value problems with complex microparticle geometry more accurately and more efficiently than previous approaches, such as ensemble averaging and homogenization.
\end{abstract}

\begin{document}


	
\maketitle
\vspace{0.25\baselineskip}
\section{Introduction}

How do we simulate the electrostatic potential due to ions around a bilipid membrane (\cref{fig:Membrane})? Or the concentration of ozone forming in clouds due to the photochemical effect (\cref{fig:smoke})? These are natural phenomena where we know how to model the underlying physical processes mathematically, using partial differential equations (PDEs) with appropriate boundary conditions. Yet despite continued advances---in computer graphics and other sciences---on numerical PDE solvers, both phenomena remain challenging to simulate \emph{accurately} and \emph{efficiently}. The root of the challenge is the same in both phenomena: Even though the problem geometry at the macroscopic level is relatively simple---we can accurately model the outer shell of the cloud in \cref{fig:smoke}, or the lipids and proteins in \cref{fig:Membrane}, with good efficiency---at the microscopic level the geometry becomes extremely complex---we need to specify the position of every water droplet forming the cloud, or every ion around the membrane, a daunting task. These are but two examples of a much broader class of natural phenomena that involve similarly complex \emph{microparticle geometry}, and thus are likewise challenging to simulate. This class includes liquid flow in porous or colloidal media, diffusive processes across biology, or heat transfer in granular media.

Our goal in this paper is to develop numerical methods that enable accurate and efficient simulation of PDEs in the presence of complex microparticle geometry, as in the above examples. Obviously, we are not the first to consider this problem. Perhaps most prominent among previous approaches are \emph{homogenization methods} \citep{marchenko2008homogenization}, which deal with complex microparticle geometry by modeling only the \emph{asymptotic} behavior of the PDE solution, for example, as particles become infinitesimally small and infinitely dense. Such limiting cases effectively eliminate (``homogenize'') the microparticle geometry, and are thus tractable to simulate. Unfortunately, the resulting PDE solutions are not always good approximations of the true behavior: Real problem settings never exactly match the homogenization assumptions, and the degree of deviation from these assumptions can vary drastically across the domain of each individual problem, making it difficult to provide accuracy guarantees or means to control the approximation error.

The need to develop numerical methods that can deal with complex microparticle geometry also arises in other areas of computer graphics. For example, in rendering, such methods are needed to simulate photorealistic images of scenes that include materials such as clouds, smoke, fog, and biological tissue. Accurate simulation of light transport in such scenes requires accounting for the multiple interactions of light with the microparticle geometry of these materials, thus posing the same computational challenge as in PDE simulation. To overcome this challenge, the predominant approach in rendering has been to represent materials with complex microparticle geometry as \emph{participating media}: Rather than enumerate explicit configurations of microscopic particles, such media model the microparticle geometry \emph{stochastically}, through bulk statistical properties such as the average particle density at different parts of the material. Combining this representation with Monte Carlo rendering methods---such as path tracing---has led to the development of \emph{volume rendering} algorithms  \citep{novak2018monte}---such as volumetric path tracing---as a highly successful methodology for the simulation of light transport in materials with complex microparticle geometry. In particular, volume rendering algorithms compute the average (in the sense of statistical expectation) light transport behavior inside such materials, without requiring the limiting assumptions of homogenization methods. This computational capability has facilitated applications in areas well beyond computer graphics, including remote sensing, chemistry, material science, and medicine.

In this paper, we set out to bring the same computational capability to PDE simulation in problem settings with complex microparticle geometry. To achieve this goal, we focus on \emph{Monte Carlo algorithms} for simulation of \emph{linear elliptic PDEs} \citep{sawhney2020monte}---for example \emph{walk on spheres} and \emph{walk on stars}---to leverage their striking similarity to Monte Carlo rendering algorithms, as well as other advantages they provide over alternative grid-based simulation methods. We first formally introduce the problem of PDE simulation in participating media (\cref{sec:background}), then specialize to so-called \emph{exponential media}---the most common type of participating media, assuming independent particles described statistically by the \emph{Poisson Boolean model} for stochastic geometry (\cref{sec:pbm}). 

We then develop two new algorithms, \emph{volumetric walk on spheres} (\cref{sec:method}) and \emph{volumetric walk on stars} (\cref{sec:neumann}), which generalize their non-volumetric namesakes to support simulation of linear elliptic PDEs in participating media. Our development mimics that of volume rendering algorithms for exponential media, taking advantage of the close similarities between Monte Carlo algorithms for simulation and rendering (\cref{sec:rendering}). We demonstrate the accuracy and efficiency of our algorithms (\cref{sec:evaluation}) through comparisons with homogenization and other baseline approaches (ensemble averaging). Lastly, we show example applications of our algorithms through simulations of natural phenomena with complex microparticle geometry (\cref{sec:physics})---in particular the two phenomena at the start of this section, the electrostatic potential of bilipid membranes (\cref{fig:Membrane}), and the photochemical effect in clouds (\cref{fig:smoke}). We provide an open-source implementation on the project website.%
\footnote{{\url{https://imaging.cs.cmu.edu/volumetric_walk_on_spheres}}}

\section{Related work}

Our work bridges ideas from PDE simulation, rendering, and stochastic geometry. We review related literature across these areas.


\paragraph{Monte Carlo PDE simulation} Our work continues the development of \emph{Monte Carlo algorithms} for PDE simulation, generalizing algorithms such as \emph{walk on spheres} \citep{muller1956wos} and \emph{walk on stars} \citep{sawhney2023wost,Miller:Robin:2024} to problems involving participating media. Despite their long history in applied mathematics and other areas \citep{sabelfeld2016stochastic}, Monte Carlo algorithms for PDE simulation were only recently introduced to computer graphics \citep{sawhney2020monte}. Since their introduction, they have been gaining popularity as an alternative to traditional grid-based methods (finite elements and boundary elements \citep{Hunter:2001:BEM,Costabel:1987:BEM}), thanks to the critical advantages they provide---output sensitivity, parallelism, robustness to imperfect geometry, and compatibility with varied geometric representations \citep[Section 1]{sawhney2020monte}. Graphics research in the past five years has seen the rapid development of these algorithms to support much broader types of PDEs \citep{Rioux-Lavoie:2022:MCFluid,Bati:2023:Coupling,de2023heat,sugimoto2024velocity,Sawhney:2022:VCWoS} and boundary conditions \citep{nabizadeh2021kelvin,Miller:Robin:2024,sawhney2023wost,Sugimoto:2023:WoB}, as well as improve efficiency \citep{qi2022bidirectional,li2024neural,Zilu:2023:NeuralCaches,bakbouk2023mean,BVC}.

We aim to further extend the capabilities of Monte Carlo PDE simulation algorithms, to enable simulation of problems with extremely complex microparticle geometry. Such problems arise in the modeling of natural phenomena such as flow effects in porous or colloidal media \citep{kadivar2021review}, diffusive effects in biology \citep{rothschild1992application,brydges1980debye} (\cref{fig:Membrane}), and radiative-diffusive photochemical effects in clouds \citep{faust1994photochemistry} (\cref{fig:smoke}). Simulating these phenomena has been a long-standing challenge also for grid-based simulation methods, typically necessitating the use of \emph{homogenization methods}, as we review below. We introduce a fundamentally different approach to addressing this challenge, based on representations of microparticle geometry as \emph{participating media}. Our approach is inspired by the similarity between Monte Carlo algorithms for PDE simulation and rendering, as we elaborate next.

\paragraph{Participating media and volume rendering} Participating media have a long history in computer graphics, and especially rendering  \citep{drebin1988volume}, as a methodology for tractably modeling microparticle geometry. Rather than pin down an exact particle configuration, they model \emph{stochastic} configurations where particle properties (location, shape) are random variables determined by the medium properties. Different statistical models give rise to different types of media, including \emph{exponential media} with independent spherical or anisotropic particles \citep{jakob2010anisotropy,heitz2015sggx}, or non-exponential media with correlated (e.g., repulsive or attractive) particles \citep{bitterli2018framework,jarabo2018radiative,d2018reciprocal,d2019reciprocal}.

Participating media have enabled light transport simulation in microparticle geometry of extreme complexity \citep{meng2015multi,muller2016efficient,moon2007rendering}, through tailored \emph{volume rendering} algorithms (such as volumetric path tracing) \citep{novak2018monte}. These algorithms generalize Monte Carlo rendering algorithms for deterministic geometry (such as path tracing) \citep{veach1998robust}, to account for geometry stochasticity inside participating media. This generalization requires only replacing geometric queries used by the deterministic algorithms (ray casting) with routines that instead sample random query outcomes from distributions determined by the medium properties (free-flight distance sampling). The development of efficient and accurate such routines has been a fruitful research area \citep{novak2014residual,raab2006unbiased,kutz2017spectral,kettunen2021unbiased,miller2019null,georgiev2019integral}.

Volume rendering algorithms simulate \emph{expected} light transport in participating media, and are used well beyond computer graphics, for example in remote sensing \citep{levis2017multiple,levis2015airborne,salesin2024unifying,salesin2024unifyingii}, chemistry \citep{berne2000dynamic,weitz1993diffusing}, and medical imaging \citep{alterman2021imaging,bar2019monte}. Recent work has extended volume rendering algorithms to stochastic \emph{macroscopic} geometry \citep{miller2024oav,vicini2021non,seyb2024stochastic}, for example arising due to acquisition noise or incomplete surface information \citep{sellan2022stochastic,sellan2023neural}. We aim to enable similar capabilities in PDE simulation, by developing volumetric Monte Carlo simulation algorithms. This development is aided by the structural similarity between Monte Carlo algorithms for simulation and rendering---as in rendering, all we need to do is replace geometric queries in simulation algorithms (closest point queries) with appropriate sampling routines (closest point sampling).

\paragraph{Homogenization of PDEs.} Homogenization methods provide an alternative methodology for analyzing and estimating solutions to linear elliptic PDEs in domains perforated by microparticle geometry. These methods consider the limit case as particles become infinitesimally small and at the same time infinitely dense. At that limit, the PDE solution asymptotically approaches the solution to a PDE with additional \emph{screening} in a domain without the microparticle geometry---\citet{marchenko2008homogenization} provide rigorous statements. Starting with the work of \citet{papanicolaou1980diffusion}, this asymptotic behavior has been shown to hold for different types of microparticle geometry, including periodic \citet{cioranescu1983terme} (\citet{cioranescu1997strange} provides an English translation) and stochastic with independent or correlated particles \citep{caffarelli2009random,calvo2015homogenization,giunti2018homogenization}. Analogous asymptotic results also hold for more general PDEs outside the scope of our paper, for example homogenizing the Stokes flow equation into the Darcy-Brinkman equation \citep{giunti2019homogenisation,whitaker1986flow,brinkman1949calculation}. 

Unfortunately, the solution to the homogenized PDE can be a poor approximation to the true solution for problems that are far from the homogenization limit, for example with particles of modest size or small density. Additionally, the approximation error can vary considerably at different parts of the PDE domain, for example near detailed geometry versus far from the boundary. Our volumetric method overcomes these issues by considering the expected, rather than asymptotic, PDE solution, and remains accurate across particle and density scales, and throughout the entire domain. We revisit homogenization and show experimental comparisons in \cref{sec:homogenization_comparison}. Lastly, homogenization methods have proven successful in graphics for simulation problems involving microscopic and multi-scale geometry beyond particle perforations \citep{yuan2024volumetric,desbrun2013modeling,sperl2020homogenized,kharevych2009numerical}.

\section{Background and problem statement}\label{sec:background}

We first review boundary value problems (BVPs) with linear elliptic PDEs and Dirichlet boundary conditions, and the walk on spheres algorithm for solving such BVPs. We then introduce the problem of solving BVPs in participating media---the focus of our paper.

\subsection{Notation}\label{sec:notation}

\begin{figure}[t]
	\centering
	\includegraphics[width=\linewidth]{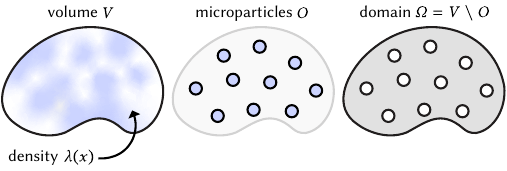}
	\caption{We consider domains $\domain \coloneq \medium \setminus \scene$ equal to the difference between a deterministic volume $\medium$ and a random configuration of particles $\scene$ contained in the volume. The particle configuration follows the Poisson Boolean model (PBM) whose density $\rate\paren{\point}$ is defined over the volume.}\label{fig:domain}
\end{figure}

We work in three dimensions and use the Euclidean norm for distances. Given any set $\queryset \subset \R^3$ and point $\point \in \R^3$, we define the functions $\closest\paren{\point,\queryset} \coloneq \argmin_{\pointalt\in\queryset}\norm{\point-\pointalt}$ and $\distop\paren{\point,\queryset} \coloneq \norm{\point-\closest\paren{\queryset,\point}}$ returning the closest point and shortest distance (resp.) to $\point$ among points in $\queryset$. We denote by $\ball\paren{\point,\radius}$ the ball with center $\point$ and radius $\radius$. For any two points $\point,\pointalt \in \R^3$, we define $\dirop\paren{\point,\pointalt}\coloneq\nicefrac{\pointalt-\point}{\norm{\pointalt-\point}}$ as the unit vector pointing from $\point$ to $\pointalt$.

Throughout the paper, we use three sets: the \emph{domain} $\domain$, the \emph{volume} $\medium$, and the \emph{microparticle geometry} $\scene$, with boundaries $\partial\domain$, $\partial\medium$, and $\partial\scene$ (resp.). We detail their roles and relationships in \cref{sec:laplace,sec:media}. For any point $\point \in \R^3$, we denote its closest points on the boundary of these sets as $\closestpoint\paren{\point} \coloneq \closest\paren{\point,\partial\domain}$, $\closestpointvol\paren{\point} \coloneq \closest\paren{\point,\partial\medium}$, and $\closestpointmicro\paren{\point} \coloneq \closest\paren{\point,\partial\scene}$; we also denote the corresponding shortest distances $\largestradius\paren{\point} \coloneq \norm{\point-\closestpoint\paren{\point}}$, $\largestradiusvol\paren{\point} \coloneq \norm{\point-\closestpointvol\paren{\point}}$, and $\largestradiusmicro\paren{\point} \coloneq \norm{\point-\closestpointmicro\paren{\point}}$. 
Given a sequence of points $\point_0, \point_1, \dots$ (e.g., a random walk), we use the shorthand notation $\closestpoint_k \coloneq \closestpoint\paren{\point_k}$ and $\largestradius_k \coloneq \largestradius\paren{\point_k}$ whenever the meaning is clear from context; we do likewise for closest points and shortest distances relative to $\partial\medium$ and $\partial\scene$, and other quantities that depend on such sequences later in the paper. 

We denote by $\Delta$ the 
Laplace operator on $\R^3$, and by $\poisson : \R_{\ge 0} \to \R$ the rotationally symmetric Poisson kernel of the zero-Dirichlet Laplace equation on a ball \citep[Appendix A]{sawhney2023wost}, parameterized with a radius $\radius$.

\subsection{The Laplace equation}\label{sec:laplace}

To simplify exposition, throughout \crefrange{sec:background}{sec:method} we consider a prototypical BVP involving the \emph{Laplace equation} with \emph{Dirichlet boundary conditions}. Our methods can extend to other linear elliptic PDEs that can be simulated with walk on spheres (e.g., screened Poisson equation). It is also possible to extend to other boundary conditions (e.g., Neumann, Robin) that can be simulated using the \emph{walk on stars} algorithm \citep{sawhney2023wost,Miller:Robin:2024}. We delay discussion of this case till \cref{sec:neumann}, and for now focus on the BVP:
\begin{equation}
	\begin{array}{rclll}
		\label{eqn:bvp_dirichlet}
		\Delta \solution\paren{\point} &=& 0 &\text{ in }& \domain, \\
		\solution\paren{\point} &=& \boundary\paren{\point} &\text{ on }& \partial\domain.
	\end{array}
\end{equation}
Here, $\domain \subset \R^3$ is the domain of the BVP, $\boundary: \partial\domain\to\R$ is the \emph{Dirichlet boundary data}, and $\solution: \domain \to \R$ is the solution we want to estimate.

\subsection{Walk on spheres}\label{sec:wos}

When the domain $\domain$ is deterministic, the \emph{walk on spheres} (WoS) algorithm \citep{sawhney2020monte, muller1956wos} computes stochastic estimates to the solution $\solution$ of the BVP \labelcref{eqn:bvp_dirichlet}, through recursive single-sample Monte Carlo integration. The starting point of the derivation of WoS is to represent the solution $\solution\paren{\point_0}$ at a point $\point_0\in\domain$ as an integral of the solution over a sphere centered at $\point_0$. A single-sample Monte Carlo estimate of this integral requires estimating the solution $\solution\paren{\point_1}$ at a point $\point_1$ sampled on the sphere. WoS estimates $\solution\paren{\point_1}$ by iterating the same single-sample Monte Carlo estimation, resulting in a recursive procedure that performs a random walk $\point_0, \point_1, \dots$. The walk terminates when it reaches a point $\point_{k}$ within a small distance $\varepsilon > 0$ from the domain boundary $\partial\domain$, where the solution is approximated by the known Dirichlet boundary data $\boundary$ at the boundary point closest to $\point_{k}$---a so-called \emph{$\varepsilon$-shell approximation}.

Concretely, we can express the solution $\solution$ to the BVP \labelcref{eqn:bvp_dirichlet} at a point $\point \in \domain$ using the \emph{boundary integral equation} (BIE) \citep{Costabel:1987:BEM}:
\begin{equation}
	\label{eqn:bie}
	\solution\paren{\point} = \int_{\sphere\paren{\point,\largestradius\paren{\point}}} \poisson(\largestradius\paren{\point})\solution\paren{\pointalt} \surfMeasure\paren{\pointalt},
\end{equation}
where $\surfMeasure$ is the surface area measure. Starting at a point $\point_0$, recursive single-sample Monte Carlo estimation of this equation with the $\varepsilon$-shell approximation results in the WoS estimator:
\begin{equation}
	\label{eqn:wos}
	\angled{\solution\paren{\point_k}} \coloneq 
		\begin{cases}
			\boundary(\closestpoint_k), &\largestradius_k < \varepsilon, \\
			\frac{\poisson(\largestradius_k)}{\pdf(\largestradius_k)}\angled{\solution\paren{\point_{k+1}}}, &\text{otherwise.}
		\end{cases}
\end{equation}
At each step, WoS performs a \emph{closest point query} to determine the boundary point $\closestpoint_k \in \partial\domain$ closest to the current walk point $\point_k$, then sets $\largestradius_k \coloneq \Vert\point_k - \closestpoint_k\Vert$. The next walk point $\point_{k+1}$ is sampled on the sphere $\partial\ball(\point_k,\largestradius_k)$ with uniform probability $\pdf(\largestradius_k) \coloneq \nicefrac{1}{4\pi(\largestradius_k)^2}$.

\subsection{Boundary value problems in participating media}\label{sec:media}

\begin{figure}[t]
	\centering
	\includegraphics[width=\linewidth]{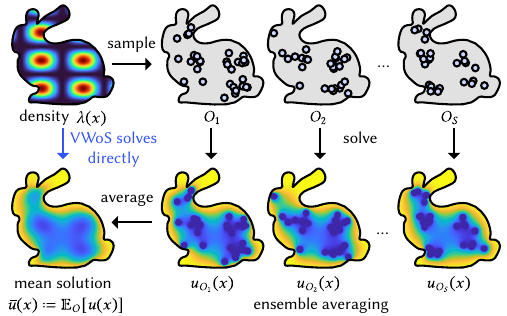}
	\caption{\emph{Ensemble averaging} is a simple but expensive method to estimate the mean solution of a PDE in a participating medium, by first sampling many random particle configurations \emph{(top row)}, then solving the PDE on each sampled domain \emph{(bottom row)}, and finally averaging the computed solutions. Our \emph{volumetric walk on spheres} algorithm directly estimates the mean solution without expensive ensemble averaging.}\label{fig:ensemble}
\end{figure}


Our focus is instances of the BVP \labelcref{eqn:bvp_dirichlet} where the domain $\domain$ is perforated by \emph{stochastic microparticle geometry}, known only up to a probabilistic model specified through the properties of a \emph{participating medium}. The stochasticity acts as a convenient abstraction to alleviate the complexity of exact modeling of microscopic geometry (for example, in domains that comprise an intractably large number of particles, such as tissue, clouds, colloidal suspensions, or porous rock formations). Before we specialize the type of participating medium in \cref{sec:pbm}, we use this section to set up such BVPs and explain the notion of solution we are interested in. 

As we show in \cref{fig:domain}, we consider the microparticle geometry $\scene\subset\R^3$ to be the union of a configuration of particles. We assume that all particles are inside a \emph{deterministic} domain $\medium \subset \R^3$ with boundary $\partial\medium$; following rendering, we term $\medium$ the \emph{participating medium} or \emph{volume}. Then, the domain of the BVP \labelcref{eqn:bvp_dirichlet} is equal to the part of the volume not occupied by particles: 
\begin{equation}\label{eqn:domain}
	\domain \coloneq \medium \setminus \scene.
\end{equation}
Additionally, as now any point in $\medium$ may be on $\partial\domain$, we extend the domain of the Dirichlet boundary data, $\boundary: \medium\to\R$. 

The configuration of particles is \emph{random}, with a distribution determined by the medium properties (e.g., particle density, \cref{sec:pbm}). Thus, the BVP solution $\solution$ is a random variable, and we want to compute its expected value, which we term the \emph{mean solution} $\meansolution$:
\begin{equation}\label{eqn:mean_solution}
	\meansolution\paren{\point} \coloneq \Exp{\scene}{\solution\paren{\point}} = \int_{\powerset\paren{\medium}} \pdf\paren{\scene} \solution\paren{\point} \ud \scene.
\end{equation}
Here, the probability $\pdf\paren{\scene}$ of a particle configuration $\scene$, the integration domain, and the integration measure depend on the stochastic microparticle geometry model, as we detail in \cref{sec:pbm}.


\paragraph{Ensemble averaging} We can estimate $\meansolution$ using \emph{ensemble averaging} (\cref{fig:ensemble}), which involves:
\begin{enumerate*}
	\item sampling many particle configurations $\scene_s$ from $\pdf$;
	\item computing the solution $\solution_{\scene_s}$ for each $\scene_s$ using, e.g., WoS;
	\item averaging the computed solutions.
\end{enumerate*}
The resulting estimate:
\begin{align}
	\angled{\meansolution\paren{\point}}_{\text{EA}} \coloneq \frac{1}{S} \sum_{s=1}^S \solution_{\scene_s}\paren{\point},
\end{align}
is consistent as $S\to \infty$ and unbiased. Though simple to implement and invaluable for validating correctness of alternative methods, ensemble averaging is impractically expensive: It requires repeatedly sampling large particle configurations $\scene_s$, and performing expensive PDE solves in the resulting complex domains. 
The high computational cost of ensemble averaging is well documented in rendering \citep{bar2019monte,bitterli2018framework}, where it has motivated the development of \emph{volume rendering} algorithms \citep{novak2018monte} that simulate light transport in participating media without ensemble averaging. These algorithms typically specialize to specific models of stochastic microparticle geometry, most commonly the \emph{Poisson Boolean model}. Motivated by the success of volume rendering algorithms, our work uses this model to develop simulation algorithms for unbiased estimation of the mean solution $\meansolution$ that elide ensemble averaging. We detail the Poisson Boolean model in \cref{sec:pbm}, then develop our algorithms in \cref{sec:method}. We elaborate on the relative merits of our algorithms compared to ensemble averaging in \cref{sec:evaluation}, where we also show experimental comparisons.

\section{Poisson Boolean model and exponential media}\label{sec:pbm}

To model the stochastic microparticle geometry in participating media, we use the \emph{Poisson Boolean model} (PBM), which is commonplace in scientific and engineering applications \citep{kadivar2021review,speidel2018topological}. In computer graphics, this model underlies volume rendering algorithms for \emph{exponential media} \citep{novak2018monte}. The widespread use of the PBM is for reasons of both modeling accuracy---it is appropriate for phenomena involving \emph{independent} particles--- and computational convenience---it is endowed with a wealth of mathematical properties that facilitate simulation.

We first provide the definition of the PBM, then explain how to use it to perform \emph{closest point sampling}, the key sampling procedure we will need in \cref{sec:method} to generalize WoS to participating media. The results we present have close analogues in volume rendering, as we detail in \cref{sec:rendering}. 
We focus on the simplest form of the PBM, which assumes that the microparticle geometry comprises \emph{spherical particles of a fixed radius}. Our presentation follows \citet[Chapters 16--17]{last2017lectures} and \citet[Chapters 2--3]{chiu2013stochastic}, and we refer to these textbooks for more detailed treatments, including generalizations to other types of particles.

\begin{dfn}[label={def:pbm}]{Poisson Boolean model}{pbm}
	We consider a function $\rate : \medium \to \R_{\ge 0}$ such that $\int_{\medium} \rate\paren{\point} \ud \point < \infty$, and a scalar $\size \in \R_{\ge 0}$. A stochastic microparticle geometry $\scene \subset \R^3$ follows the \emph{Poisson Boolean model} with \emph{density} $\rate$ and \emph{size} $\size$ if it equals a union of balls $\scene \coloneq \bigcup_{\iterball=1}^\numballs \ball\paren{\centerpoint_\iterball,\size}$ such that the set of centers $\centers_\scene \coloneq \curly{\centerpoint_\iterball \in \medium}_{\iterball=1}^\numballs$ is a \emph{Poisson point process} on $\medium$ with rate function $\rate$. Equivalently:
	\begin{itemize}[leftmargin=*]
		\item the number of balls is a Poisson-distributed integer random variable, $\numballs \sim \Pois\paren{\int_{\medium} \rate\paren{\point} \ud \point}$;
		\item conditionally on $\numballs$, the centers are independent and distributed proportionally to the density%
		\footnote{The density $\rate$ is also known as the \emph{intensity function} of the Poisson point process.}
		$\rate$,  $\pdf\paren{\conditional{\centerpoint_\iterball}{\numballs}} \propto \rate\paren{\centerpoint_\iterball}$.
	\end{itemize}
\end{dfn}
We often consider the special case of the \emph{homogeneous} PBM with constant density $\rate\paren{\point} \homoeq \rate$, %
%
%
and refer to the general case as the \emph{heterogeneous} model. We write $\scene\sim \PBM\paren{\rate, \size}$ for a particle configuration that follows the PBM with rate $\rate$ and size $\size$. 

\newcommand{\dirichletContactFigure}{%
	\setlength{\columnsep}{0.5em}
	\setlength{\intextsep}{-0.15em}
	\begin{wrapfigure}[13]{r}{0.35\linewidth}
		\centering
		\vspace{-0.25em}
		\includegraphics[width=\linewidth]{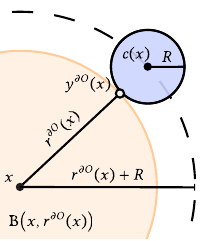}
		\vspace{-2.25em}
		\caption{Computing the closest point $\closestpointmicro\!(\point)$ from the closest center $\centerpoint(\point)$.}\label{fig:centers_points}
	\end{wrapfigure}}

\dirichletContactFigure{}\paragraph{Closest point distribution and sampling} As WoS interacts with the domain through closest point queries, using it in participating media requires reasoning about the \emph{random} closest point $\closestpointmicro\paren{\point} \in \partial\scene$ between a point $\point \in \medium$ and the boundary $\partial\scene$ of the stochastic microparticle geometry $\scene$. This random point follows the so-called \emph{closest point distribution}, whose probability density function (PDF) $\probcp_{\point}$ is known analytically when $\scene$ follows the PBM \citep[Section 16.3]{last2017lectures}. We discuss this distribution in more detail in \cref{sec:proof}; here we focus instead on \emph{sampling} from it, which we need to perform WoS in participating media (\cref{sec:method}).

As $\scene$ is a union of spherical particles, we perform closest point sampling by first sampling the \emph{closest particle center} $\closestcenter\paren{\point} \in \centers_\scene$ to $\point$. We then determine the closest point $\closestpointmicro\paren{\point}$ to $\point$, and associated shortest distance $\largestradiusmicro\paren{\point}$, from $\closestcenter\paren{\point}$ as (\cref{fig:centers_points}):
\begin{equation}\label{eqn:center_to_contact}
	\largestradiusmicro\paren{\point} = \norm{\point - \closestcenter\paren{\point}}-\size,\quad
	\closestpointmicro\paren{\point} = \point + \largestradiusmicro\paren{\point} \dirop\paren{\point,\closestcenter\paren{\point}}.
\end{equation}
To sample $\closestcenter\paren{\point}$, we use the \emph{polar representation} of the Poisson point process describing $\centers_\scene$ \citep[Section 7.4]{last2017lectures}.

\begin{prp}[label={pro:polar}]{Polar representation of\\ Poisson point processes}{prp}
	We assume that the set of centers $\centers_\scene \coloneq \curly{\centerpoint_\iterball \in \medium}_{\iterball=1}^\numballs$ is a Poisson point process on $\medium$ with rate function $\rate$. For any point $\point \in \medium$, we let $\closestcenter\paren{\point} \in \centers_\scene$ be its random closest center, and $\largestradiuscenter\paren{\point} \coloneq \norm{\point - \closestcenter\paren{\point}}$ the random shortest distance-to-center. Then:
	\begin{enumerate}[leftmargin=*]
		\item The shortest distance-to-center $\largestradiuscenter\paren{\point}$ has PDF:
		\begin{equation}\label{eqn:pdc}
			\probdc_{\point}\paren{\radius} \coloneq \exp\paren{-\rateset\paren{\point,\radius}} \int_{\partial\ball\paren{\point,\radius}} \rate\paren{\pointalt} \surfMeasure\paren{\pointalt},
		\end{equation}
		where we define:
		\begin{equation}
			\rateset\paren{\point, \radius} \coloneq \int_{\ball\paren{\point,\radius}} \rate\paren{\pointalt} \ud \pointalt. \label{eqn:rate_pbm}
		\end{equation}
		\item The closest center $\closestcenter\paren{\point}$ has \emph{conditional} PDF given $\largestradiuscenter\paren{\point} = \radius$:
		\begin{equation}\label{eqn:pcc_cond}
			\probcc_{\point}\paren{\conditional{\pointalt}{\radius}} \coloneq \frac{\rate\paren{\pointalt}}{\int_{\partial\ball\paren{\point,\radius}} \rate\paren{\pointalt} \surfMeasure\paren{\pointalt}},
		\end{equation}
		for $\pointalt \in \partial\ball\paren{\point, \radius}$, and zero everywhere else.%
		\footnote{More precisely, the PDF is a Dirac delta on $\partial\ball\paren{\point,\radius}$ with respect to the area measure.} 
	\end{enumerate}
\end{prp}
In the homogeneous case, \cref{eqn:pdc,eqn:pcc_cond} simplify to:
\begin{align}
	\probdc_{\point}\paren{\radius} &\homoeq \exp\paren{-\nicefrac{4}{3}\pi\paren{\radius}^3\rate} 4 \pi \paren\radius^2 \rate, \label{eqn:pdc_homo} \\
	\probcc_{\point}\paren{\conditional{\pointalt}{\radius}} &\homoeq \frac{1}{4 \pi \radius^2}. \label{eqn:pcc_cond_homo}
\end{align}
As we detail in \cref{sec:rendering}, the PDF $\probdc_{\point}$ is analogous to the \emph{free-flight distribution} in volume rendering of exponential media \citep{novak2018monte,bitterli2018framework}.

\begin{figure}[t]
	\centering
	\includegraphics[width=\linewidth]{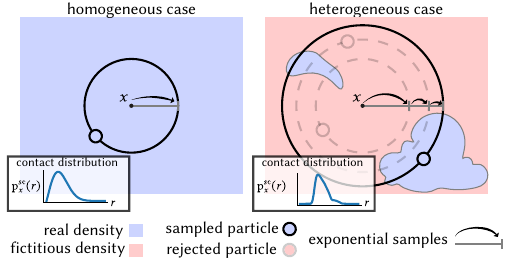}
	\caption{The Poisson Boolean model enables efficient closest point sampling---or equivalently closest particle center sampling (\cref{fig:centers_points}). In a homogeneous medium (left), we first sample an exponential random variable to determine the (cubed) distance to the closest center, then uniformly sample the center itself on a sphere of appropriate radius. In a heterogeneous medium (right), we use thinning to first sample multiple centers---in order of increasing distance---from a medium homogenized through the addition of fictitious density, randomly accept or reject the sampled centers, then use the first accepted one as the sampled closest center.}\label{fig:cpsampling}
\end{figure}

\Cref{eqn:pdc} (or \labelcref{eqn:pdc_homo} in the homogeneous case) implies that the \emph{cubed distance} $\paren{\largestradiuscenter\paren{\point}}^3$ from $\point$ to $\closestcenter\paren{\point}$ is an \emph{exponential random variable} with rate $\rateset\paren{\ball\paren{\point,\radius}}$. This property and \cref{eqn:pcc_cond} (or \labelcref{eqn:pcc_cond_homo} in the homogeneous case) allow sampling $\closestcenter\paren{\point}$ by first sampling its cubed distance from $\point$, then sampling a point on the corresponding sphere around $\point$ \citep[Section 2.5]{chiu2013stochastic} (\cref{fig:cpsampling}):
\begin{itemize}[leftmargin=*]
	\item In the homogeneous case, we first sample an exponential random variate $\xi \sim \Expo\bracket{\nicefrac{4}{3}\pi\rate}$, equal to the cubed distance from $\point$ to $\closestcenter\paren{\point}$. We then sample $\closestcenter\paren{\point}$ uniformly on $\partial\ball(\point, \cbrt{\xi})$.
	\item In the heterogeneous case, we use the \emph{thinning} method for sampling heterogeneous Poisson processes \citep{lewis1979simulation}. Given a \emph{majorant density} $\majrate \ge \rate\paren{\point},\,\forall \point \in \R^3$, we sequentially sample exponential random variates $\xi_1,\xi_2,\dots \sim \Expo\bracket{\nicefrac{4}{3}\pi\majrate}$. For each $\xi_s$, we sample a point $\centerpoint_s$ uniformly on $\partial\ball(\point, \cbrt{\sum_{t=1}^s \xi_t})$, and randomly accept or reject it with acceptance probability $\nicefrac{\rate\paren{\centerpoint_s}}{\majrate}$. Then, we set $\closestcenter\paren{\point}$ as the first accepted point $\centerpoint_s$.
\end{itemize}
After sampling $\closestcenter\paren{\point}$, we set $\largestradiusmicro\paren{\point}$ and $\closestpointmicro\paren{\point}$ using \cref{eqn:center_to_contact}. This procedure can return $\largestradiusmicro\paren{\point} < 0$, corresponding to a case where $\point$ is inside a particle, $\point \in \scene$---it will be convenient for the algorithm we develop in \cref{sec:derivation} to distinguish this case by setting $\closestpointmicro\paren{\point} \coloneq \point$. We summarize the sampling procedure for the heterogeneous case in \cref{alg:closest_sampling}, which reduces to that for the homogeneous case when $\rate$ is constant and we use $\majrate \coloneq \rate$.

Intuitively, we can interpret \cref{alg:closest_sampling} as follows: We fill the volume $\medium$ with \emph{fictitious density} until the sum of real and fictitious density equals $\majrate$ everywhere. We then sample particles from the resulting homogeneous medium using exponential sampling, in order of increasing distance from $\point$. We reject particles due to the fictitious density until we find the first particle due to the real density. As we detail in \cref{sec:rendering}, \cref{alg:closest_sampling} is analogous to the \emph{delta tracking} algorithm in volume rendering for sampling free-flight distances in heterogeneous exponential media \citep{coleman1968mathematical,raab2006unbiased}, which has a similar intuitive interpretation in terms of fictitious density \citep{miller2019null,novak2014residual}. 

\begin{algorithm}[t]
	\caption{Closest point sampling in the Poisson Boolean model.}\label{alg:closest_sampling}
	\begin{algorithmic}[1]
		\algblockdefx[Name]{Struct}{EndStruct}
		[1][Unknown]{\textbf{struct} #1}
		{}
	\algtext*{EndStruct}
	\algblockdefx[Name]{FORDO}{ENDFORDO}
		[1][Unknown]{\textbf{for} #1 \textbf{do}}
		{}
	\algtext*{ENDFORDO}
	\algblockdefx[Name]{WHILEDO}{ENDWHILEDO}
		[1][Unknown]{\textbf{while} #1 \textbf{do}}
		{}
	\algtext*{ENDWHILEDO}
	\algblockdefx[Name]{IF}{ENDIF}
		[1][Unknown]{\textbf{if} #1 \textbf{then}}
		{}
	\algtext*{ENDIF}
	\algblockdefx[Name]{ELSE}{ENDELSE}
		{\textbf{else}}
		{}
	\algtext*{ENDELSE}
	\algblockdefx[Name]{IFTHEN}{ENDIFTHEN}
		[2][Unknown]{\textbf{if} #1 \textbf{then} #2}
		{}
	\algtext*{ENDIFTHEN}
	\algblockdefx[Name]{RETURN}{ENDRETURN}
		[1][Unknown]{\textbf{return} #1}
		{}
	\algtext*{ENDRETURN}
	\algblockdefx[Name]{COMMENT}{ENDCOMMENT}
		[1][Unknown]{\textcolor{commentcolor}{\(\triangleright\) \textit{#1}}}
		{}
	\algtext*{ENDCOMMENT}

	\Require \textit{A query point $\point$, a majorant density $\majrate$, a struct implementing the PBM density $\rate$, the PBM particle size $\size$.}
 	\Ensure \textit{Closest point $\closestpointmicro$.}
	\Function{SampleClosestPoint}{$\point,\ \majrate,\ \rate,\ \size$}
	 	\COMMENT[Initialize cubed radius]\ENDCOMMENT
		\State $\cubedradius \gets 0$
		\WHILEDO[\textsc{true}]
			\COMMENT[Sample exponential variate]\ENDCOMMENT
			\State $\xi \gets \Proc{SampleExponential}(\nicefrac{4}{3}\pi\majrate)$
			\COMMENT[Increment cubed radius]\ENDCOMMENT
			\State $\cubedradius \mathrel{+}= \xi$
			\COMMENT[Uniformly sample a point on the unit sphere]\ENDCOMMENT
			\State $\directionalt \gets \Proc{SampleUnitSphere()}$
			\COMMENT[Compute particle center]\ENDCOMMENT
			\State $\centerpoint \gets \point + \cbrt{\cubedradius}\ \directionalt$
			\COMMENT[Compute acceptance probability]\ENDCOMMENT
			\State $\alpha \gets \rate.\Proc{Evaluate}(\centerpoint)\ /\ \majrate$
			\COMMENT[Accept or reject the particle center]\ENDCOMMENT
			\IFTHEN[$\alpha > \Proc{SampleUniform(0, 1)}$]{\textbf{break}}
			\ENDIFTHEN
		\ENDWHILEDO
		\COMMENT[Compute distance to closest point]\ENDCOMMENT
		\State $\largestradiusmicro \gets \Vert\point - \centerpoint\Vert - \size$
		\COMMENT[Check if $\point$ is inside the sampled particle]\ENDCOMMENT
		\IFTHEN[$\largestradiusmicro < 0$]{\textbf{return} $\point$}
			\ENDIFTHEN
		\COMMENT[Compute sampled closest point]\ENDCOMMENT
		\State $\closestpointmicro \gets \point + \largestradiusmicro\ \dirop(\point,\centerpoint)$
		\RETURN[$\closestpointmicro$]\ENDRETURN
	\EndFunction
	\end{algorithmic}
\end{algorithm}

\paragraph{Conditional closest point sampling.} Our derivation in \cref{sec:method} additionally requires sampling the closest point $\closestpointmicro\paren{\point} \in \partial\scene$ \emph{conditionally} on events of the form $\queryset\cap \scene = \emptyset$ for various sets $\queryset\subset\medium$---i.e., knowing that no point of the microparticle geometry is in $\emptyset$. We prove in \cref{sec:proof} that we can do so using \cref{alg:closest_sampling}, after replacing $\rate$ with the \emph{conditional density} $\rate(\conditionals{\cdot}{\queryset}) : \medium \to \R_{\ge 0}$:
\begin{equation}\label{eqn:conditional_rate}
	\rate(\conditionals{\point}{\queryset}) \coloneq
	\begin{cases}
		0, & \point \in \queryset^{\oplus\size}, \\
		\rate(\point), &\text{otherwise},
	\end{cases}
\end{equation}
where $\queryset^{\oplus\size}\coloneq \curly{\point \in \R^3 : \distop\paren{\point, \queryset} \le \size}$ is the dilation of $\queryset$ by a ball of radius $\size$. Intuitively, zeroing out the density inside $\queryset^{\oplus\size}$ ensures that centers sampled closer to $\queryset$ than $\size$ will be rejected, thus guaranteeing $\queryset \cap \scene = \emptyset$.

\section{Volumetric walk on spheres}\label{sec:method}

\begin{figure}[t]
	\centering
	\includegraphics[width=\linewidth]{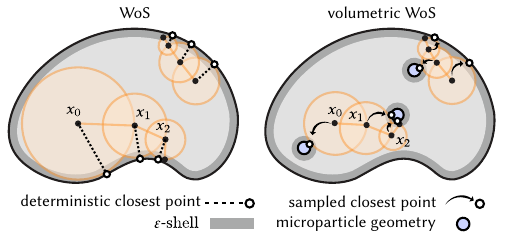}
	\caption{(\emph{Left}) Walk on spheres jumps along spheres whose radius it determines through closest point queries with respect to the deterministic domain boundary. It terminates when it reaches the $\varepsilon$-shell of the boundary. (\emph{Right}) Volumetric walk on spheres also jumps along spheres, but determines their radius through closest point \emph{sampling} with respect to both stochastic microparticle geometry and the deterministic domain. It terminates when it reaches the $\varepsilon$-shell of either the microparticle geometry or the domain.}\label{fig:walks}
\end{figure}

We now derive a recursive Monte Carlo estimator for the mean solution $\meansolution$ of the BVP \labelcref{eqn:bvp_dirichlet} in exponential participating media. Our estimator generalizes the standard WoS estimator \labelcref{eqn:wos}, by leveraging one of its critical properties \citep[Section 1]{sawhney2020monte}: WoS interacts with the BVP domain $\domain$ only through closest point queries to determine, at each point $\point$, its 
closest point $\closestpoint$ on $\partial\domain$. Thus, when $\domain$ includes stochastic microparticle geometry, it suffices to replace these closest point queries with closest point sampling from appropriate distributions (\cref{fig:walks}). The resulting \emph{volumetric walk on spheres} (VWoS) estimator (\cref{eqn:vwos} and \cref{alg:vwos}) enables direct simulation of the mean solution without ensemble averaging, while remaining structurally very close to WoS.

\subsection{Derivation}\label{sec:derivation}

At a high level, our derivation of VWoS mimics that of WoS: We first derive an integral equation (\cref{eqn:vbie}) for the mean solution $\meansolution$---an analogue of the BIE \cref{eqn:bie}---then apply recursive single-sample Monte Carlo estimation. There is, however, an important difference with WoS: As we recurse, the integral equation changes, to \emph{condition} on the history accumulated during previous steps. To build intuition about this conditioning, it is convenient to separately derive the initial and subsequent steps of the VWoS recursion. 

\paragraph{Initial step.} To evaluate the mean solution $\meansolution$ at an initial point $\point_0 \in \medium$, we use expectation on both sides of the BIE \labelcref{eqn:bie}:
\begin{equation}\label{eqn:deriv1}
	\meansolution\paren{\point_0} = \Exp{\scene}{\int_{\sphere(\point_0,\largestradius_0)}\poisson(\largestradius_0)\solution\paren{\point_1} \surfMeasure\paren{\point_1}}.
\end{equation}
Besides $\solution$, the only random quantity in the right-hand-side integral is the distance $\largestradius_0 \coloneq \norm{\point_0 - \closestpoint_0}$ between $\point_0$ and $\closestpoint_0$, the random point on $\partial\domain$ closest to $\point_0$. We use the shorthand $\probcp_0 \coloneq \probcp_{\point_0}$ for the PDF of $\closestpoint_0$---the closest point distribution in \cref{sec:pbm} modified to account for the deterministic medium boundary $\partial\medium$. Using the law of total expectation, we rewrite \cref{eqn:deriv1} as:
\begin{align}\label{eqn:vbie_init}
	\meansolution\paren{\point_0} = \int_{\medium} \probcp_0(\cpvar_0) \int_{\sphere(\point_0,\distvar_0)} \!\!\!\!\!\!\!\poisson(\distvar_0)\meansolution(\conditionals{\point_1}{\closestpoint_0 = \cpvar_0}) \surfMeasure\paren{\point_1} \ud \cpvar_0,
\end{align}
where $\distvar_0 \coloneq \norm{\point_0 - \cpvar_0}$ and $\conditionals{\cdot}{\cdot}$ is probabilistic conditioning. 

To derive an estimator for $\meansolution\paren{\point_0}$, we first use single-sample Monte Carlo for both integrals in \cref{eqn:vbie_init}. 
\begin{enumerate}
	\item We sample a point $\cpvar_0 \sim \probcp_0$ by: using \cref{alg:closest_sampling} to sample a closest point $\closestpointmicro_0$ on the stochastic $\partial\scene$; using a closest point query to determine the closest point $\closestpointvol_0$ on the deterministic $\partial\medium$; then setting $\cpvar_0 \coloneq \closest(\point_0, \{\closestpointmicro_0,\closestpointvol_0\})$.
	\item We sample a point $\point_1$ uniformly on the sphere $\partial\ball(\point_0,\radius_0)$.
\end{enumerate}
How we proceed depends on the sampled distance $\distvar_0$. If $\distvar_0$ is smaller than a threshold $\varepsilon$, we use an $\varepsilon$-shell approximation to set $\meansolution\paren{\point_0}$ equal to the Dirichlet boundary data $\boundary\paren{\point_0}$, and terminate.%
\footnote{This $\varepsilon$-shell approximation differs from that in WoS (\cref{eqn:wos}): It evaluates $\boundary$ directly at $\point_0$, which is possible due to the extension of $\boundary$ to all of $\medium$. In practice, for small particle sizes $\size$, the two approximations behave very similarly.}
Importantly, this approximation also covers the case where $\distvar_0 = 0$, which means that \cref{alg:closest_sampling} determined that $\point_0$ is inside a particle (\cref{sec:pbm}). Otherwise, we need to recursively estimate $\meansolution(\conditionals{\point_1}{\closestpoint_0 = \cpvar_0})$, as we explain below. Thus, we arrive at the estimator:
\begin{equation}
	\label{eqn:vwos_init}
	\langle\meansolution(\point_0)\rangle \coloneq 
		\begin{cases}
			\boundary(\point_0), &\radius_0 < \varepsilon, \\
			\frac{\poisson(\radius_0)}{\pdf(\radius_0)}\langle\meansolution(\conditionals{\point_1}{\closestpoint_0 = \cpvar_0})\rangle, &\text{otherwise}.
		\end{cases}
\end{equation}

\paragraph{Subsequent steps.} We now consider the $k$-th step of the VWoS recursion. After consecutive applications of the law of total expectation and single-sample Monte Carlo---analogously to \cref{eqn:deriv1,eqn:vwos_init} (resp.)---we must \emph{condition} on closest points on the domain boundary sampled in all previous steps. Intuitively, each sampled closest point pins down part of the stochastic microparticle geometry that subsequent closest point sampling procedures must remember and continue to respect. To simplify notation, we define the walk \emph{memory} $\memory_k$ accumulated at step $k$ as the logical conjunction of closest point sampling outcomes in previous steps:
\begin{equation}\label{eqn:memory}
	\memory_k \coloneq \closestpoint_0 = \cpvar_0 \wedge \dots \wedge \closestpoint_{k-1} = \cpvar_{k-1}, \quad \memory_0 \coloneq \emptyset.
\end{equation}
We elaborate on the interpretation of $\memory_k$, and how to use it for conditional closest point sampling, in \cref{sec:conditional_sampling}.

Then, at the $k$-th step of the VWoS recursion, we must estimate the conditional mean solution $\meansolution(\conditionals{\point_k}{\memory_k})$. Exactly analogously to the case $k=0$ and \cref{eqn:vbie_init}, we use the BIE \labelcref{eqn:bie} and the law of total expectation to derive an integral equation for $\meansolution(\conditionals{\point_k}{\memory_k})$, which we term the \emph{boundary integral equation in participating media}.

\begin{myTitledBox}{Boundary integral equation in participating media}
	At the $k$-step of a random walk with memory $\memory_k$ as in \cref{eqn:memory}, the conditional mean solution $\meansolution$ to the Laplace equation of \labelcref{eqn:bvp_dirichlet} in a participating medium $\medium$ satisfies:
	\begin{align}\label{eqn:vbie}
		&\meansolution\paren{\conditional{\point_k}{\memory_k}} = \int_{\medium}\probcp_k(\conditionals{\cpvar_k}{\memory_k}) \nonumber \\
		&\quad\quad\cdot\int_{\sphere(\point_k,\distvar_k)} \poisson(\distvar_k)\meansolution(\conditionals{\point_{k+1}}{\memory_{k+1}}) \surfMeasure\paren{\point_{k+1}} \ud \cpvar_k,
	\end{align}
	where $\distvar_k \coloneq \norm{\point_k - \cpvar_k}$.
\end{myTitledBox}

Exactly analogously to the derivation of the estimator \labelcref{eqn:vwos_init} for \cref{eqn:vbie_init}, we use single-sample Monte Carlo and the $\varepsilon$-shell approximation to derive an estimator for \cref{eqn:vbie}, which we term the \emph{volumetric walk on spheres} (VWoS) estimator.
\begin{myTitledBox}{Volumetric walk on spheres estimator}
	A recursive single-sample Monte Carlo estimator for \cref{eqn:vbie} is given by:
	\begin{equation}
		\label{eqn:vwos}
		\langle\meansolution(\conditionals{\point_k}{\memory_k})\rangle \coloneq 
			\begin{cases}
				\boundary(\point_k), &\radius_k < \varepsilon, \\
				\frac{\poisson(\radius_k)}{\pdf(\radius_k)}\langle\meansolution(\conditionals{\point_{k+1}}{\memory_{k+1}})\rangle, &\text{otherwise,}
			\end{cases}
	\end{equation}
	where: the memories $\memory_k$ and $\memory_{k+1}$ are defined in \cref{eqn:memory}; $\radius_k \coloneq \norm{\point_k-\cpvar_k}$ with $\cpvar_k$ sampled from $\probcp_k(\conditionals{\cdot}{\memory_k})$; and the next point $\point_{k+1}$ is sampled uniformly on $\partial\ball(\point_k,\radius_k)$. 
\end{myTitledBox}
Using the definition of $\memory_0$ in \cref{eqn:memory}, \cref{eqn:vbie,eqn:vwos} subsume \cref{eqn:vbie_init,eqn:vwos_init} (resp.) by setting $k \coloneq 0$.

\cref{alg:vwos} summarizes an implementation of VWoS, highlighting differences from WoS (\cref{eqn:wos}). The two estimators are structurally near-identical, with only two differences:
\begin{itemize}[leftmargin=*]
	\item Whereas WoS determines sphere radii using deterministic closest point queries, VWoS does so using conditional closest point sampling (\cref{alg:vwos}, \emph{line \labelcref{alg:vwos:sample}}).
	\item Whereas WoS is \emph{memoryless}---each step is independent of previous steps---VWoS has full \emph{memory}---each step depends on all previous steps (\cref{alg:vwos}, \emph{lines \labelcref{alg:vwos:mem1,alg:vwos:mem2}}).
\end{itemize}
Thanks to its close similarity to WoS, VWoS maintains the advantages of WoS \citep[Section 1]{sawhney2020monte}, and is easy to implement within existing WoS libraries \citep{Sawhney:2023:Zombie}, requiring only an implementation of memory (including associated sampling and updating procedures), which we discuss next.

\begin{algorithm}[t]
	\caption{The volumetric walk on spheres estimator.\newline \textbf{Note:} Comments in \textcolor{commentorangecolor}{orange} highlight changes to walk on spheres.}
	\label{alg:vwos}
	\begin{algorithmic}[1]
		\algblockdefx[Name]{Struct}{EndStruct}
		[1][Unknown]{\textbf{struct} #1}
		{}
	\algtext*{EndStruct}
	\algblockdefx[Name]{FORDO}{ENDFORDO}
		[1][Unknown]{\textbf{for} #1 \textbf{do}}
		{}
	\algtext*{ENDFORDO}
	\algblockdefx[Name]{WHILEDO}{ENDWHILEDO}
		[1][Unknown]{\textbf{while} #1 \textbf{do}}
		{}
	\algtext*{ENDWHILEDO}
	\algblockdefx[Name]{IF}{ENDIF}
		[1][Unknown]{\textbf{if} #1 \textbf{then}}
		{}
	\algtext*{ENDIF}
	\algblockdefx[Name]{ELSE}{ENDELSE}
		{\textbf{else}}
		{}
	\algtext*{ENDELSE}
	\algblockdefx[Name]{IFTHEN}{ENDIFTHEN}
		[2][Unknown]{\textbf{if} #1 \textbf{then} #2}
		{}
	\algtext*{ENDIFTHEN}
	\algblockdefx[Name]{RETURN}{ENDRETURN}
		[1][Unknown]{\textbf{return} #1}
		{}
	\algtext*{ENDRETURN}
	\algblockdefx[Name]{COMMENT}{ENDCOMMENT}
		[1][Unknown]{\textcolor{commentcolor}{\(\triangleright\) \textit{#1}}}
		{}
	\algtext*{ENDCOMMENT}
	\algblockdefx[Name]{HLCOMMENT}{ENDHLCOMMENT}
		[1][Unknown]{\textcolor{commentorangecolor}{\(\triangleright\) \textit{#1}}}
		{}
	\algtext*{ENDHLCOMMENT}
	\algblockdefx[Name]{SUBCOMMENT}{ENDSUBCOMMENT}
		[1][Unknown]{\textcolor{commentorangecolor}{\(\triangleright\quad\) \textit{#1}}}
		{}
	\algtext*{ENDSUBCOMMENT}

	\Require \textit{A query point $\point$, a parameter $\varepsilon$, a majorant density $\majrate$, a struct implementing the PBM density $\rate$, the PBM particle size $\size$.}
 	\Ensure \textit{A single-sample estimate of the mean solution $\meansolution\paren{\point}$.}
	\Function{InitializeEstimation}{$\point,\varepsilon,\majrate,\rate,\size$}
		\COMMENT[Initialize empty memory]\ENDCOMMENT
		\State $\memory \gets \Proc{Memory}.\Proc{Initialize}()$
		\RETURN[$\Proc{VolumetricWalkOnSpheres}(\point, \varepsilon,\majrate,\rate,\size,\memory)$]\ENDRETURN
	\EndFunction

	\Function{VolumetricWalkOnSpheres}{$\point, \varepsilon,\majrate,\rate,\size,\memory$}
		\COMMENT[Sample closest point \textcolor{commentorangecolor}{conditionally on current memory}]\ENDCOMMENT
		\State $\cpvar \gets \Proc{SampleClosestPointWithMemory}(\point,\majrate,\rate,\size,\memory)$ \label{alg:vwos:sample}
		\COMMENT[Compute radius of next walk sphere]\ENDCOMMENT
		\State $\distvar \gets \Vert\point - \cpvar\Vert$
		
		\COMMENT[Check for $\varepsilon$-shell approximation]\ENDCOMMENT
		\IFTHEN[$\distvar < \varepsilon$]{\textbf{return} $\boundary(\point)$}
		\ENDIFTHEN

		\HLCOMMENT[Update memory:]\ENDHLCOMMENT
		\State $\memory.\Proc{Update}(\point, \cpvar)$\label{alg:vwos:mem1}
		
		\COMMENT[Uniformly sample a point on the unit sphere]\ENDCOMMENT
		\State $\directionalt \gets \Proc{SampleUnitSphere()}$
		
		\COMMENT[Compute next walk point]\ENDCOMMENT
		\State $\point \gets \point + \distvar \directionalt$

		\COMMENT[Continue from next walk point \textcolor{commentorangecolor}{with updated memory}]\ENDCOMMENT
		\RETURN[$\Proc{VolumetricWalkOnSpheres}(\point, \varepsilon,\majrate,\rate,\size,\memory)$]\label{alg:vwos:mem2}
		\ENDRETURN

	\EndFunction
	\end{algorithmic}
\end{algorithm}

\subsection{Closest point sampling with memory}\label{sec:conditional_sampling}

Intuitively, closest point sampling with memory $\memory_k$ requires respecting empty space (inside walk spheres) and sampled particles accumulated during previous walk steps. We first formalize this intuition about conditioning on $\memory_k$, then present procedures for memory updating and closest point sampling with memory in VWoS.

\begin{figure}[t]
	\centering
	{\includegraphics[width=\linewidth]{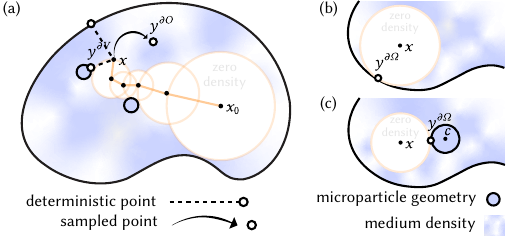}
	\phantomsubcaption\label{fig:a}
	\phantomsubcaption\label{fig:b}
	\phantomsubcaption\label{fig:c}}
	\caption{(\protect\subref{fig:a}) To sample the closest point $\closestpoint$ at $\point$ conditionally on the memory $\memory$ accumulated during a walk, we determine two points:
	%
		First, we sample the random closest point $\closestpointmicro$ on the stochastic microparticle geometry, but with the PBM density zeroed out inside the spheres formed during the walk. 
		Second, we query the closest point $\closestpointvol$ on the deterministic boundary of the medium and previously sampled particles. 
	%
	Then, we select the closest of these two points to $\point$, $\closestpoint \coloneq \closest(\point, \{\closestpointmicro, \closestpointvol\})$. After sampling, we add to $\memory$ a new empty sphere (\protect\subref{fig:b}), and a new particle if $\closestpoint$ was not on the boundary of the medium or previously sampled particles (\protect\subref{fig:c}).}\label{fig:memory}
\end{figure}

\paragraph{Understanding conditioning on memory.} As $\memory_k$ is a conjunction of multiple \emph{closest point events} (\cref{eqn:memory}), we first consider each individual such event. Conditioning on $\closestpoint_l = \cpvar_l$, for any $l = 0,\dots,k-1$, has two implications (\cref{fig:memory}(b, c)):
\begin{enumerate}[label={{C\arabic*}.},ref={{C\arabic*}}]
	\item \label{enu:C1} No point of the microparticle geometry $\scene$ is closer to walk point $\point_l$ than $\distvar_l = \norm{\point_l - \cpvar_l}$, i.e., the ball $\ball\paren{\point_l, \distvar_l}$ is empty.
	\item \label{enu:C2} If $\cpvar_l$ is not on the medium boundary $\partial\medium$, then it is on the stochastic boundary $\partial\scene$, i.e., the microparticle geometry $\scene$ includes a particle $\ball\paren{\centerpoint_l, \size}$ centered at $\centerpoint_l \coloneq \cpvar_l + \size \dirop\paren{\point_l,\cpvar_l}$. 
\end{enumerate}
Formally, from \labelcref{enu:C1}--\labelcref{enu:C2}, the closest point event $\closestpoint_l = \cpvar_l$ is equivalent to the event $\ball\paren{\point_l, \distvar_l}\cap\scene = \emptyset \wedge \centerpoint_l \in \centers_\scene$.

Next we consider the full memory $\memory_k$. It is convenient to associate with $\memory_k$ two sets summarizing the information from all closest point events it includes. We define the \emph{empty-ball memory} $\emptyspace\paren{\memory_k}$ and \emph{sampled-particle memory} $\particles\paren{\memory_k}$ as the unions of the empty balls (\labelcref{enu:C1}) and sampled particles (\labelcref{enu:C2}) (resp.) for all eligible events in $\memory_k$:
\begin{equation}\label{eqn:memories}
	\emptyspace\paren{\memory_k} \coloneq \bigcup\nolimits_{l=0}^{k-1}\ball\paren{\point_l,\distvar_l},\quad\particles\paren{\memory_k} \coloneq \bigcup\nolimits_{l: \cpvar_l\notin\partial\medium} \ball\paren{\centerpoint_l, \size}.
\end{equation}
$\particles\paren{\memory_k}$ will typically include fewer than $k$ (and maybe even zero) particles, as only steps where the sampled closest point is not on $\partial\medium$ contribute a particle (\labelcref{enu:C2}). Additionally, $\particles\paren{\memory_k}$ may include the same particle multiple times: at each step $l\in\curly{1,k}$, the closest point $\cpvar_l$ may be on the boundary of a particle sampled at a prior step $l^\prime < l$, and fixed as deterministic geometry for subsequent steps.

Using $\emptyspace\paren{\memory_k}$ and $\particles\paren{\memory_k}$, we can express the implications of conditioning on $\memory_k$ as direct generalizations of \labelcref{enu:C1}--\labelcref{enu:C2} (\cref{fig:memory}(a)): 
\begin{enumerate}[label={{M\arabic*}.},ref={{M\arabic*}}]
	\item \label{enu:M1} The space $\emptyspace\paren{\memory_k}$ is empty.
	\item \label{enu:M2} The microparticle geometry $\scene$ includes $\particles\paren{\memory_k}$.
\end{enumerate}
Formally, from \labelcref{enu:M1}--\labelcref{enu:M2}, the memory $\memory_k$ is equivalent to the event $\emptyspace\paren{\memory_k}\cap\scene = \emptyset \wedge \curly{\centerpoint_l: \cpvar_l\notin\partial\medium} \subset \centers_\scene$.

In practice, we implement memory as in \cref{alg:memory}:%
\footnote{For efficiency, this implementation does not allow duplicate particles in $\particles\paren{\memory_k}$.}
We use two list data structures that contain the center-radius pairs $(\point_l,\distvar_l)$ and particle centers $\centerpoint_l$ in \cref{eqn:memories}, along with procedures for memory updates at each walk step (\cref{alg:memory}, \emph{line \labelcref{alg:memory:update}}), containment queries on $\emptyspace\paren{\memory_k}$ (\cref{alg:memory}, \emph{line \labelcref{alg:memory:containment}}), and closest point queries on $\partial\particles\paren{\memory_k}$ (\cref{alg:memory}, \emph{line \labelcref{alg:memory:cpq}})---we use the two queries for closest point sampling with memory, as we discuss next.

\paragraph{Sampling procedure} From the above discussion, sampling a closest point $\cpvar_k$ conditionally on $\memory_k$ requires that we do not sample in the empty space $\emptyspace\paren{\memory_k}$ (\labelcref{enu:M1}), and that we consider the boundary of previously sampled particles $\particles\paren{\memory_k}$ as \emph{deterministic} (\labelcref{enu:M2}). Realizing both requirements algorithmically is straightforward. First, we use thinning on the density of the medium to remove the empty space, and augment its geometry to include the sampled particles:
\begin{align}
	\rate_k\paren{\point} \coloneq \rate\paren{\conditionals{\point}{\emptyspace\paren{\memory_k}}},\quad
	\partial\medium_k \coloneq \partial\paren{\medium \setminus \particles\paren{\memory_k}}.\label{eqn:medium_k}
\end{align}
Second, we sample $\cpvar_k \sim \probcp_k\paren{\conditionals{\cdot}{\memory_k}}$ by: using \cref{alg:closest_sampling} with $\rate_k$ to sample a closest point $\closestpointmicro_k$ on the stochastic boundary $\partial\scene$; using a closest point query to determine the closest point $\closestpointvol_k$ on the deterministic boundary $\partial\medium_k$; then setting $\cpvar_k \coloneq \closest(\point_k, \{\closestpointmicro_k,\closestpointvol_k\})$. \Cref{alg:closest_sampling_memory} summarizes this procedure. Used with empty memory ($\memory \coloneq \emptyset$), this procedure is equivalent to how we sample $\cpvar_0 \sim \probcp_0\paren{\cdot}$ at the initial step of VWoS (\cref{eqn:vwos_init}).

\begin{algorithm}[t]
	\caption{Implementation of memory.}
	\label{alg:memory}
	\begin{algorithmic}[1]
	\algblockdefx[Name]{Struct}{EndStruct}
		[1][Unknown]{\textbf{struct} #1}
		{}
	\algtext*{EndStruct}
	\algblockdefx[Name]{FORDO}{ENDFORDO}
		[1][Unknown]{\textbf{for} #1 \textbf{do}}
		{}
	\algtext*{ENDFORDO}
	\algblockdefx[Name]{WHILEDO}{ENDWHILEDO}
		[1][Unknown]{\textbf{while} #1 \textbf{do}}
		{}
	\algtext*{ENDWHILEDO}
	\algblockdefx[Name]{IF}{ENDIF}
		[1][Unknown]{\textbf{if} #1 \textbf{then}}
		{}
	\algtext*{ENDIF}
	\algblockdefx[Name]{ELSE}{ENDELSE}
		{\textbf{else}}
		{}
	\algtext*{ENDELSE}
	\algblockdefx[Name]{IFTHEN}{ENDIFTHEN}
		[2][Unknown]{\textbf{if} #1 \textbf{then} #2}
		{}
	\algtext*{ENDIFTHEN}
	\algblockdefx[Name]{RETURN}{ENDRETURN}
		[1][Unknown]{\textbf{return} #1}
		{}
	\algtext*{ENDRETURN}
	\algblockdefx[Name]{COMMENT}{ENDCOMMENT}
		[1][Unknown]{\textcolor{commentcolor}{\(\triangleright\) \textit{#1}}}
		{}
	\algtext*{ENDCOMMENT}

	\Struct[$\Proc{Memory}$]
		\State \textbf{attribute} $\emptylist$	\Comment{List of center-radius pairs for empty balls}
		\State \textbf{attribute} $\particlelist$	\Comment{List of centers of sampled particles}

		\COMMENT[Initialize empty memory]\ENDCOMMENT
		\Function{Initialize}{\vphantom{0}}
			\State $\emptylist, \particlelist \gets \emptyset$
		\EndFunction

		\COMMENT[Add an empty ball of center $\point$ and radius $\distvar$]\ENDCOMMENT
		\Function{AddEmptySphere}{$\point, \distvar$}	
			\State $\emptylist \gets \emptylist \cup \{(\point, \distvar)\}$
		\EndFunction

		\COMMENT[Add a sampled particle of center $\centerpoint$]\ENDCOMMENT
		\Function{AddSampledParticle}{$\centerpoint$}
			\IFTHEN[$\centerpoint^\prime \textbf{ in } \particlelist$]{\textbf{return}}\Comment{Skip if $\centerpoint$ already in list}\ENDIFTHEN
			\State $\particlelist \gets \particlelist \cup \{(\centerpoint)\}$
		\EndFunction

		\COMMENT[Update memory given walk point $\point$ and its closest point $\cpvar$]\ENDCOMMENT\label{alg:memory:update}
		\Function{Update}{$\point, \cpvar$}
		\COMMENT[Add walk step to list of empty balls]\ENDCOMMENT
		\State $\Proc{AddEmptySphere}(\point, \norm{\point-\cpvar})$
		\COMMENT[Add sampled particle if $\cpvar$ is not on medium boundary]\ENDCOMMENT
		\IFTHEN[$\cpvar\notin \partial\medium$]{$\Proc{AddSampledParticle}(\cpvar + \size\ \dirop(\point,\cpvar))$}\ENDIFTHEN
		\EndFunction

		\COMMENT[Perform a containment query on dilated empty space]\ENDCOMMENT
		\Function{IsInsideDilatedEmptySpheres}{$\point, \size$}\label{alg:memory:containment}
			\FORDO[$(\point^\prime,\distvar)\ \textbf{in}\ \emptylist$]
				\IFTHEN[$\Vert\point - \point^\prime\Vert < \distvar + \size$]{\textbf{return} \textsc{true}}
				\ENDIFTHEN
			\ENDFORDO
			\RETURN[\textsc{false}]\ENDRETURN
		\EndFunction
		
		\COMMENT[Perform a closest point query on sampled particle boundary]\ENDCOMMENT
		\Function{GetClosestPoint}{$\point$}\label{alg:memory:cpq}
			\COMMENT[Query closest particle center]\ENDCOMMENT
			\State $\currentnn \gets \Proc{NearestNeighbor}(\point, \particlelist)$ 
			\COMMENT[Return corresponding closest point on particle boundary]\ENDCOMMENT
			\RETURN[$\currentnn - \size\ \dirop(\point,\currentnn)$]\ENDRETURN
		\EndFunction
	\EndStruct
	\end{algorithmic}
\end{algorithm}

\begin{algorithm}[t]
	\caption{Closest point sampling with memory.}
	\label{alg:closest_sampling_memory}
	\begin{algorithmic}[1]
		\algblockdefx[Name]{Struct}{EndStruct}
		[1][Unknown]{\textbf{struct} #1}
		{}
	\algtext*{EndStruct}
	\algblockdefx[Name]{FORDO}{ENDFORDO}
		[1][Unknown]{\textbf{for} #1 \textbf{do}}
		{}
	\algtext*{ENDFORDO}
	\algblockdefx[Name]{WHILEDO}{ENDWHILEDO}
		[1][Unknown]{\textbf{while} #1 \textbf{do}}
		{}
	\algtext*{ENDWHILEDO}
	\algblockdefx[Name]{IF}{ENDIF}
		[1][Unknown]{\textbf{if} #1 \textbf{then}}
		{}
	\algtext*{ENDIF}
	\algblockdefx[Name]{ELSE}{ENDELSE}
		{\textbf{else}}
		{}
	\algtext*{ENDELSE}
	\algblockdefx[Name]{IFTHEN}{ENDIFTHEN}
		[2][Unknown]{\textbf{if} #1 \textbf{then} #2}
		{}
	\algtext*{ENDIFTHEN}
	\algblockdefx[Name]{RETURN}{ENDRETURN}
		[1][Unknown]{\textbf{return} #1}
		{}
	\algtext*{ENDRETURN}
	\algblockdefx[Name]{COMMENT}{ENDCOMMENT}
		[1][Unknown]{\textcolor{commentcolor}{\(\triangleright\) \textit{#1}}}
		{}
	\algtext*{ENDCOMMENT}

	\COMMENT[Implementation of conditional density in \cref{eqn:medium_k}]\ENDCOMMENT
	\Struct[$\Proc{ConditionalDensity}$]
		\State \textbf{attributes} $\rate,\size, \memory$ \Comment{PBM parameters and memory}
		\Function{Initialize}{$\rate^\prime, \size^\prime, \memory^\prime$}	\Comment{Initialize attributes}
			\State $\rate \gets \rate^\prime$, $\size \gets \size^\prime$, $\memory \gets \memory^\prime$
		\EndFunction
		\COMMENT[Evaluate conditional density at query point $\point$]\ENDCOMMENT
		\Function{Evaluate}{$\point$}
			\IFTHEN[$\memory.\Proc{IsInsideDilatedEmptySpheres}(\point,\size)$]
				\RETURN[$0$	\Comment{$\point$ is inside empty space}]\ENDRETURN
			\ENDIFTHEN
			\ELSE{\textbf{ return} $\rate.\Proc{Evaluate}(\point)$}\ENDELSE
		\EndFunction
	\EndStruct

	\COMMENT[Implementation of updated medium geometry in \cref{eqn:medium_k}]\ENDCOMMENT
	\Struct[$\Proc{UpdatedMediumGeometry}$]
		\State \textbf{attributes} $\partial\medium,\memory$ \Comment{Medium geometry and memory}
		\Function{Initialize}{$\partial\medium^\prime, \memory^\prime$}	\Comment{Initialize attributes}
			\State $\partial\medium \gets \partial\medium^\prime$, $\memory \gets \memory^\prime$
		\EndFunction
		\COMMENT[Perform a closest point query on updated medium boundary]\ENDCOMMENT
		\Function{GetClosestPoint}{$\point$}
			\COMMENT[Query closest points on medium and particle boundaries]\ENDCOMMENT
			\State $\cpvar \gets \partial\medium.\Proc{GetClosestPoint}(\point)$
			\State $\cpvar^\prime \gets \memory.\Proc{GetClosestPoint}(\point)$
			\RETURN[$\closest(\point, \{\cpvar, \cpvar^\prime\})$] \Comment{Return closest of two points}\ENDRETURN
		\EndFunction
	\EndStruct

	\Require \textit{A query point $\point$, a majorant density $\majrate$, a struct implementing the PBM density $\rate$, the PBM particle size $\size$, a struct implementing memory $\memory$.}
 	\Ensure \textit{Closest point $\cpvar$.}
	\Function{SampleClosestPointWithMemory}{$\point,\majrate,\rate,\size,\memory$}
		\COMMENT[Create conditional density]\ENDCOMMENT
		\State $\condrate \gets \Proc{ConditionalDensity}.\Proc{Initialize}(\rate, \size, \memory)$
		\COMMENT[Create updated medium geometry]\ENDCOMMENT
		\State $\condmedium \gets \Proc{UpdatedMediumGeometry}.\Proc{Initialize}(\partial\medium, \memory)$
		\COMMENT[Sample a closest point conditioning on empty balls]\ENDCOMMENT
		\State $\cpvar^{\partial\scene} \gets \Proc{SampleClosestPoint}(\point, \majrate, \condrate, \size)$
		\COMMENT[Query closest point on updated deterministic boundary]\ENDCOMMENT
		$\closestpointvol \gets \condmedium.\Proc{GetClosestPoint}(\point)$
		\COMMENT[Return the closest of sampled and deterministic closest points]\ENDCOMMENT
		\State $\cpvar \gets \closest(\point, \{\closestpointmicro, \closestpointvol\})$
		\RETURN[$\cpvar$]\ENDRETURN
	\EndFunction
	\end{algorithmic}
\end{algorithm}

\section{Volumetric walk on stars} \label{sec:neumann}

We now turn our attention to a generalization of the BVP \labelcref{eqn:bvp_dirichlet} that prescribes mixed Dirichlet and \emph{Neumann} boundary conditions. Our discussion is brief, as most of the concepts we introduced in \cref{sec:background,sec:method} for the Dirichlet-only case extend to this case.

\subsection{Boundary value problem}\label{sec:bvp_neumann}

We focus on estimating solutions to the BVP:
\begin{equation}
	\begin{array}{rclll}
		\label{eqn:bvp_neumann}
		\Delta \solution\paren{\point} &=& 0 &\text{ in }& \domain, \\
		\solution\paren{\point} &=& \boundary\paren{\point} &\text{ on }& \dbound, \\
		\frac{\partial\solution}{\partial\normal}\paren{\point} &=& 0 &\text{ on }& \nbound.
	\end{array}
\end{equation}
Compared to \cref{eqn:bvp_dirichlet}, here we partition $\partial\domain$ into a subset $\dbound$ where the solution has prescribed values $\solution$ (Dirichlet boundary conditions), and a subset $\nbound$ where the solution has prescribed normal derivatives $\nicefrac{\partial\solution}{\partial\normal}$ (Neumann boundary conditions).

\subsection{Walk on stars}\label{sec:wost}

When the domain $\domain$ is deterministic, the \emph{walk on stars} (WoSt) algorithm \citep{sawhney2023wost,Miller:Robin:2024} estimates the BVP solution $\solution$ in a manner analogous to WoS---recursive single-sample Monte Carlo estimation of an appropriate BIE. In particular, instead of \cref{eqn:bie} for the Dirichlet-only problem, WoSt starts from the following BIE for the mixed Dirichlet-Neumann problem \labelcref{eqn:bvp_neumann}:
\begin{equation}\label{eqn:bie_neumann}
	\solution\paren{\point} = \int_{\partial\stardomain\paren{\point,\starradius\paren{\point}}} \poisson(\starradius\paren{\point})\solution\paren{\pointalt} \surfMeasure\paren{\pointalt}.
\end{equation}
The integration domain is the boundary of a \emph{star-shaped region} $\stardomain(\point,\starradius\paren{\point})$ defined as follows \citep[Section 4]{sawhney2023wost}: For any point $\point \in \domain$, we denote by $\dpoint\paren{\point} \coloneq \closest(\point, \dbound)$ its closest point on $\dbound$, and by $\npoint\paren{\point}$ its closest point on the \emph{visibility silhouette} of $\nbound$. We denote by $\dradius\paren{\point} \coloneq \Vert\point - \dpoint\paren{\point}\Vert$, $\nradius\paren{\point} \coloneq \Vert\point - \npoint\paren{\point}\Vert$ the distances to these points, and by $\starradius\paren{\point} \coloneq \min\{\dradius\paren{\point},\nradius\paren{\point}\}$ their minimum. Then, the star-shaped region equals $\stardomain(\point,\starradius\paren{\point}) \coloneq \ball(\point,\starradius\paren{\point}) \cap \domain$. Importantly, $\partial\stardomain(\point,\starradius\paren{\point})$ can include parts of $\nbound$ but not $\dbound$---except, potentially, for the closest point $\dpoint\paren{\point}$ when $\starradius\paren{\point} = \dradius\paren{\point}$.

Using recursive single-sample Monte Carlo estimation on \cref{eqn:bie_neumann}, WoSt performs a random walk $\point_0,\point_1,\dots$ that terminates using the same $\varepsilon$-shell approximation as in WoS. At each walk point $\point_k$, the expression for the WoSt estimator is the same as in \cref{eqn:wos}, except replacing $\largestradius_k$ with $\starradius_k \coloneq \starradius\paren{\point_k}$. At each walk step, WoSt performs a closest point query to determine $\dpoint_k \coloneq \dpoint\paren{\point_k}$, a \emph{closest silhouette point query} to determine $\npoint_k \coloneq \npoint\paren{\point_k}$, and directional sampling to determine the next walk point $\point_{k+1}$.

\subsection{Participating media}

We next consider the mixed Dirichlet-Neumann BVP \labelcref{eqn:bvp_neumann} when the domain $\domain$ includes stochastic microparticle geometry. We follow the setup of \cref{sec:media}, defining $\domain$ as in \cref{eqn:domain}, and assume that $\scene \sim \PBM(\rate,\size)$. We further assume that the Dirichlet boundary coincides with the deterministic medium boundary, and the Neumann boundary comprises the boundary of the particles; that is: 
\begin{equation}\label{eqn:dirichlet_neumann}
	\dbound \coloneq \partial\medium\ \text{ and }\ \nbound \coloneq \partial\scene. 
\end{equation}
We choose this problem specification to simplify exposition, but we can extend our method to the case where $\dbound$ and $\nbound$ each include both deterministic and stochastic boundaries.

Given the close similarity between \cref{eqn:bie,eqn:bie_neumann}, and between WoS and WoSt, we can adapt the derivation in \cref{sec:derivation} exactly analogously to the mixed Dirichlet-Neumann BVP. The result is a \emph{volumetric walk on stars} (VWoSt) estimator that has the same form as the VWoS estimator \labelcref{eqn:vwos}, but with $\distvar_k$ being the radius used to form the star-shaped region $\stardomain\paren{\point_k,\distvar_k}$. As explained in \cref{sec:wost}, $\distvar_k$ equals the minimum of $\Vert\point_k - \dpoint_k\Vert$ and $\Vert\point_k - \npoint_k\Vert$. From \cref{eqn:dirichlet_neumann}, $\dpoint_k$ is the \emph{deterministic} closest point $\closestpointvol_k$ on $\partial\medium$, and $\npoint_k$ is the \emph{random} closest silhouette point on $\partial\scene$. VWoSt determines $\npoint_k$ through \emph{closest silhouette point sampling}, conditional on the memory $\memory_k$ of all such points sampled in previous steps. 

\newcommand{\neumannContactFigure}{%
	\setlength{\columnsep}{0.5em}
	\setlength{\intextsep}{-0.15em}
	\begin{wrapfigure}[11]{r}{0.35\linewidth}
		\centering
		\includegraphics[width=\linewidth]{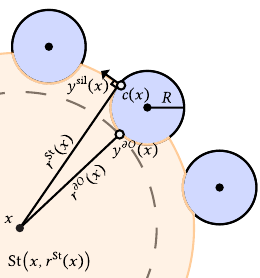}
		\vspace{-2.25em}
		\caption{Sampling to form a star-shaped region.}\label{fig:stars}
	\end{wrapfigure}}

\subsection{Closest silhouette point sampling with memory}\label{sec:conditional_silhouette_sampling}

As the microparticle geometry $\scene$ comprises spherical particles of the same radius $\size$, the closest silhouette point on $\partial\scene$ will lie on the same particle as the closest point on $\partial\scene$ (\cref{fig:stars}). Thus, we perform closest point sampling with memory as in \cref{sec:conditional_sampling} to determine the closest point $\closestpointmicro_k$, then compute from it $\npoint_k$ and $\nradius_k$ analytically. From these values, we compute the radius $\distvar_k$ for the star-shaped region $\stardomain(\point_k,\distvar_k) = \ball(\point_k, \distvar_k) \cap \scene$.

\neumannContactFigure{}
Because $\distvar_k$ can be greater than the shortest distance $\largestradiusmicro_k = \Vert\point_k - \closestpointmicro_k\Vert$ to $\partial\scene$, forming the star-shaped region $\stardomain(\point_k,\distvar_k)$ requires also determining any additional particles that are closer to $\point_k$ than $\distvar_k$. We do so by using \cref{alg:closest_sampling} with conditional density $\rate(\conditionals{\cdot}{\emptyspace(\memory_k)})$ (\cref{eqn:medium_k}) to continue sampling particles beyond the closest one, in order of increasing distance from $\point_k$, until we exceed the distance $\distvar_k$. As these particles become fixed geometry for subsequent steps, we include them in the memory $\memory_k$---and in particular, in the sampled-particle memory $\particles\paren{\memory_k}$ (\cref{eqn:memories}).

\section{Experimental evaluation}\label{sec:evaluation}

Our experimental evaluation includes: 
\begin{enumerate*}
	\item comparisons of VWoS and VWoSt with ensemble averaging to both validate unbiasedness and assess performance (\cref{sec:homogenization_comparison}); 
	\item comparisons with homogenization methods (\cref{sec:homogenization_comparison}); and 
	\item analysis of the impact of memory (\cref{sec:memory_ablation}). 
\end{enumerate*}
Our experiments demonstrate that VWoS and VWoSt provide performance and accuracy benefits over both ensemble averaging and homogenization, though the magnitude of these benefits depends on the experimental setup. 

\paragraph{Implementation details} We implement VWoS and VWoSt in \textsc{Zombie}~\citep{Sawhney:2023:Zombie}, making only minor modifications to its WoS and WoSt routines. Our implementation supports mixed boundary conditions on the medium boundary $\partial\medium$, and both Dirichlet and Neumann conditions on the microparticle geometry boundary $\partial\scene$. We represent density as either an analytic function or a dense grid. When using analytic functions, we compute a majorant for the largest deterministic empty ball at each point; when using a grid, we compute a global majorant. We set the $\varepsilon$-shell parameter to be at least one order of magnitude smaller than the particle radius. \Cref{tab:scene_parameters,tab:scene_boundary_conditions} provide details (e.g., medium properties, boundary conditions, algorithmic parameters) for each experiment.

We implement ensemble averaging also in \textsc{Zombie}, with the following optimizations:
\begin{enumerate*}
	\item We share sampled particle configurations across evaluation points, to amortize sampling costs.
	\item We construct an, also shared, bounding value hierarchy (BVH) for each configuration, to accelerate closest point queries.
\end{enumerate*}
For the BVH, we use \textsc{FCPW} \citep{Sawhney:2021:FCPW} and extend it to support spherical primitives.

\subsection{Comparison to ensemble averaging}\label{sec:ensemble_averaging}

\begin{figure*}[t]
	\centering
	{\includegraphics{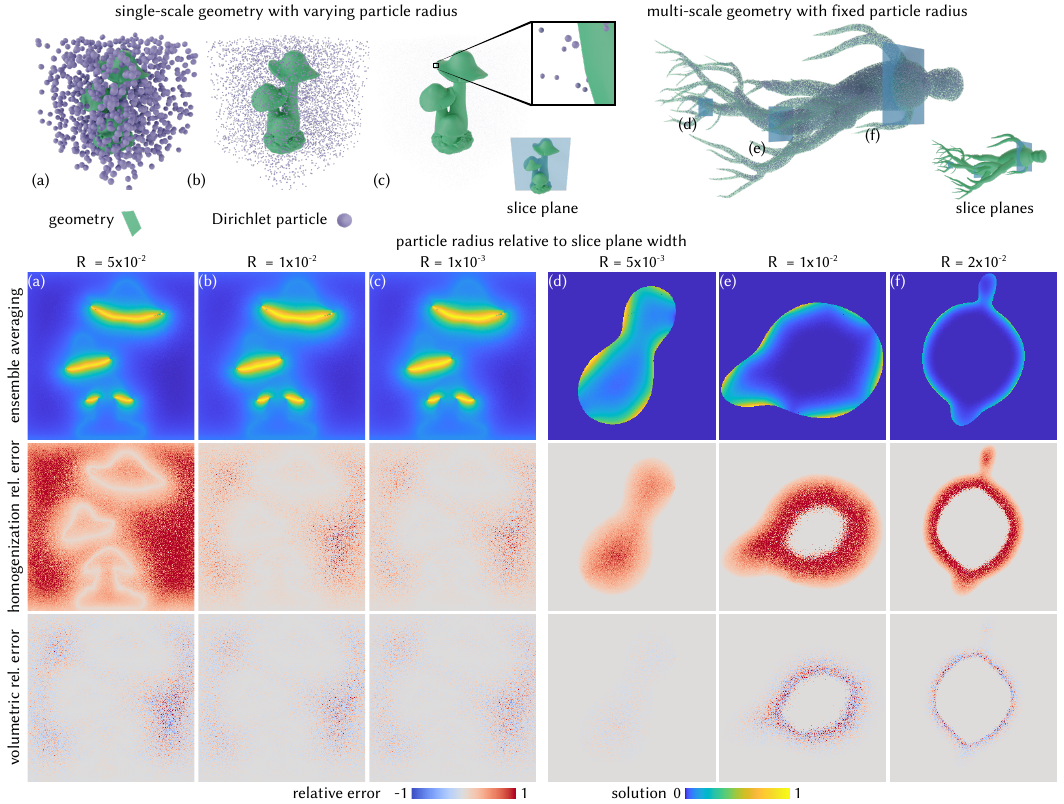}
	\phantomsubcaption\label{fig:mushroom_a}
	\phantomsubcaption\label{fig:mushroom_b}
	\phantomsubcaption\label{fig:mushroom_c}
	\phantomsubcaption\label{fig:ginseng_d}
	\phantomsubcaption\label{fig:ginseng_e}
	\phantomsubcaption\label{fig:ginseng_f}}
	\caption{We compare the outputs of VWoS (\emph{fourth row}) and homogenization (\emph{third row}) to the reference mean solution (\emph{second row}) computed with ensemble averaging, in BVPs with Dirichlet-only boundary conditions. Across a range of medium parameters and boundaries, VWoS reliably produces unbiased mean solution estimates, whereas homogenization introduces noticeable bias as the particle size increases (\protect\subref{fig:mushroom_a}, \protect\subref{fig:mushroom_b}), or at geometrically thin parts of the volume (\protect\subref{fig:ginseng_d}, \protect\subref{fig:ginseng_e}). For each experiment, we visualize (\emph{first row}) a representative sampled configuration of the microparticle geometry.}
	\label{fig:homogenization_comparison}
\end{figure*}

We compare our VWoS and VWoSt algorithms against ensemble averaging, in BVPs involving various volume geometries and medium parameters. These comparisons aim to both validate the consistency and unbiasedness of our algorithms---by ensuring that their mean solution estimates match those from ensemble averaging---and quantify the performance improvements they provide.

\paragraph{Volumetric walk on spheres.} \Cref{fig:homogenization_comparison} shows comparisons between VWoS and ensemble averaging, in BVPs with Dirichlet-only boundary conditions. In the mushroom domain of \cref{fig:homogenization_comparison}(\subref{fig:mushroom_a}--\subref{fig:mushroom_c}), we vary particle size and density across nearly two orders of magnitude. In the ginseng root domain of \cref{fig:homogenization_comparison}(\subref{fig:ginseng_d}--\subref{fig:ginseng_f}), we evaluate the solution near geometric features whose scale varies by a factor of $4$. 

In all experiments, VWoS estimates match the mean solution estimates from ensemble averaging. Moreover, VWoS is more than $3\times$ faster than ensemble averaging in terms of the time it takes to perform the same number of walks---and thus reach the same variance---when accounting for both particle configuration sampling and solve time in ensemble averaging. This performance difference becomes more stark in experiments requiring sparse evaluation points or dense media; we revisit this point in \cref{sec:membrane} where, depending on the density of evaluation points, VWoS is one-to-several orders of magnitude faster than ensemble averaging. The slow performance of ensemble averaging is because of three reasons:
\begin{enumerate}
	\item Ensemble averaging has poor output sensitivity, as it must sample entire particle configurations even to compute the solution at only one point. Thus, ensemble averaging wastes considerable compute sampling particles in large parts of the volume that have little to no impact to its output.
	\item Ensemble averaging has poor sample amortization, as it must sample a new particle configuration for each walk, or at least batch of walks. Sharing sampled particle configurations across evaluation points mitigates this issue only partially.
	\item Ensemble averaging has poor geometric query performance, as it must perform closest point queries against entire particle configurations. Even with logarithmic-complexity query implementations (e.g., with a BVH), large particle numbers in dense media introduce considerable computational overhead.
\end{enumerate}
By contrast, in VWoS, each walk samples few particles, on demand, and only in its locality; the result is good output sensitivity, no sample waste, and no overhead in geometric queries. Thus, VWoS greatly improves performance relative to ensemble averaging, analogous to the performance improvements volume rendering algorithms provide over ensemble averaging for light transport simulation.

\begin{figure}[t]
	\centering
	{\includegraphics[width=\columnwidth]{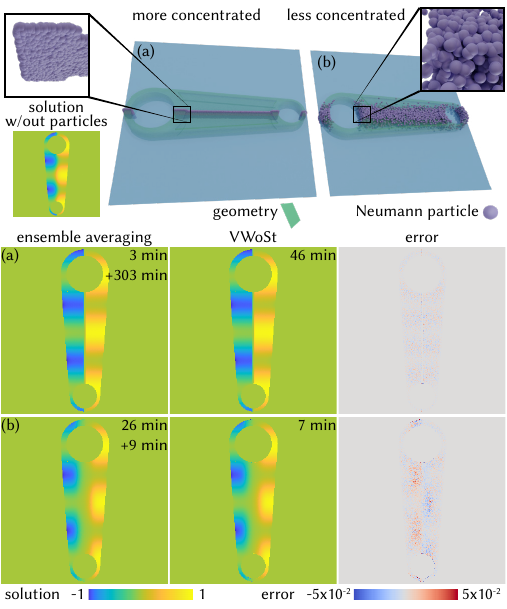}
	\phantomsubcaption\label{fig:more_conc_a}
	\phantomsubcaption\label{fig:less_conc_b}}
	\caption{We compare the output of VWoSt (\emph{second column}) to the reference mean solution (\emph{first column}) computed with ensemble averaging, in BVPs with Neumann boundary conditions on the microparticle geometry. We also report the runtimes of VWoSt and ensemble averaging (the $+$ numbers are the time to sample particle configurations for ensemble averaging). We experiment with more concentrated (\emph{second row}) and less concentrated (\emph{third row}) particle densities. The error images (\emph{third column}) show that VWoSt correctly estimates the mean solution nearly $5\times$ faster than ensemble averaging. For each experiment, we visualize (\emph{first row}) a representative sampled configuration of the microparticle geometry.}
	\label{fig:neumann_validation}
\end{figure}

\paragraph{Volumetric walk on stars} \Cref{fig:neumann_validation} shows comparisons between VWoSt and ensemble averaging, in BVPs with mixed Dirichlet-Neumann boundary conditions. We perform experiments on the same scene using two densities, which are Gaussian-shaped along one dimension and constant along the other two. 

In both cases, VWoSt matches the estimates of ensemble averaging and improves performance (measured as in the VWoS experiments above). In \cref{fig:neumann_validation}(\subref{fig:more_conc_a}), the concentrated density impacts both methods: it reduces configuration sampling efficiency in ensemble averaging, and requires large majorants for closest silhouette point sampling in VWoSt. 
In \cref{fig:neumann_validation}(\subref{fig:less_conc_b}), where density is less concentrated, sampling configurations in ensemble averaging is much cheaper. However, walks are more expensive in ensemble averaging than in VWoSt, because they perform closest silhouette point queries against entire configurations. 


\subsection{Comparison to homogenization}\label{sec:homogenization_comparison}

\emph{Homogenization methods}~\citep{giunti2018homogenization} transform a BVP with stochastic microparticle geometry into a \emph{homogenized BVP} that involves a modified PDE in deterministic geometry. The solution $\homogenizedsolution$ of the homogenized BVP converges to the mean solution $\meansolution$ of the original BVP only \emph{asymptotically} at the limit of infinitesimally small and infinitely dense particles ($\size\to 0$, $\rate\to\infty$, while $\rate\size$ remains constant). For the BVP \labelcref{eqn:bvp_dirichlet}, the homogenized BVP is:%
\footnote{This homogenization procedure is distinct from using a homogeneous participating medium to perform closest point sampling (\cref{alg:closest_sampling}).}
\begin{equation}
	\begin{array}{rclll}
		\label{eq:homogenizedpde}
		\Delta \homogenizedsolution\paren{\point} - 4 \pi \rate \size \homogenizedsolution\paren{\point} &= & 0 &\text{ in }& \medium,\\
		\homogenizedsolution\paren{\point} &=& \boundary\paren{\point} &\text{ on }& \partial\medium.
	\end{array}
\end{equation}
The homogenized BVP can then be solved efficiently using standard WoS. However, for any finite values of $\size$ and $\rate$, homogenization provides \emph{biased} estimates of the mean solution, with bias increasing as $\size$ or $\rate$ deviate more from the asymptotic case. 
Bias can also vary at different domain points, e.g., near geometrically thin versus thick parts. By contrast, VWoS is \emph{unbiased} and provides accurate estimates \emph{robustly} across particle properties and domain points.

We demonstrate these advantages experimentally in \cref{fig:homogenization_comparison}. In the mushroom domain \cref{fig:homogenization_comparison}(\subref{fig:mushroom_a}--\subref{fig:mushroom_c}), increasing the particle size leads to noticeable bias in the homogenization solution. In the ginseng root domain \cref{fig:homogenization_comparison}(\subref{fig:ginseng_d}--\subref{fig:ginseng_f}), for a fixed particle radius, bias is higher at thin parts of the root (where the particle radius is comparable to the domain size) than at thick parts (where the particle radius is much smaller than the domain size). Additionally, bias is higher near the medium boundary than away from it. Thus, even though homogenization is $2--20\times$ faster than VWoS in these experiments, the solution estimates it produces have considerable bias that is hard to control across problem settings and complex domains.

\subsection{Analysis of impact from memory}\label{sec:memory_ablation}

\begin{figure}[t]
	\centering
	\includegraphics[width=\columnwidth]{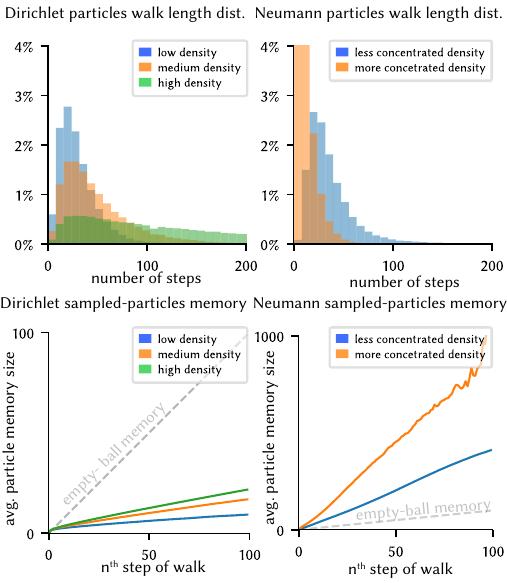}
	\caption{Statistics for walk length (\emph{first row}) and memory size (\emph{second row}) for the mushroom (\protect\cref{fig:homogenization_comparison}(\subref{fig:mushroom_a}--\subref{fig:mushroom_c}), Dirichlet boundary conditions) and connector (\protect\cref{fig:neumann_validation}(\subref{fig:more_conc_a}--\subref{fig:less_conc_b}), Neumann boundary conditions) domains. Though the size of the empty-ball memory always equals walk length, the size of the sampled-particle memory can grow slower (in the Dirichlet case) or faster (in the Neumann case) than walk length. In both cases, increased density leads to longer walks and faster growth of sampled-particle memory.}
	\label{fig:memory_statistics}
\end{figure}

\begin{figure}[t]
	\centering
	{\includegraphics[width=\columnwidth]{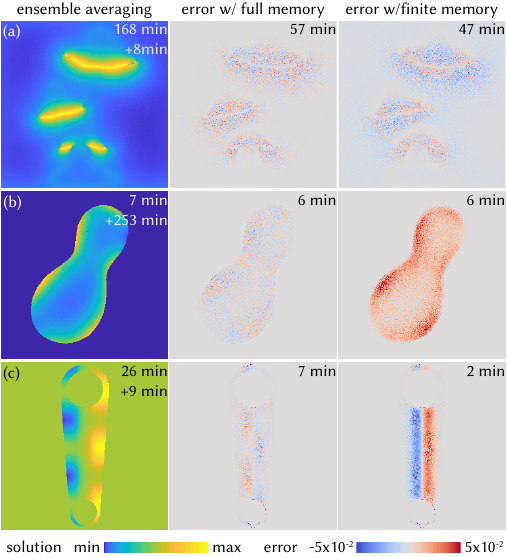}
	\phantomsubcaption\label{fig:memory_mushroom_a}
	\phantomsubcaption\label{fig:memory_ginseng_b}
	\phantomsubcaption\label{fig:memory_connector_c}}
	\caption{We compare the bias-performance trade-off of finite memory of size one (\emph{third column}) versus full memory (\emph{second column}) using ensemble averaging (\emph{first column}) for reference. We show error images and report runtimes for each method (the + numbers are the time to sample particle configurations for ensemble averaging). Using finite memory improves runtime only marginally (\protect\subref{fig:memory_mushroom_a}, \protect\subref{fig:memory_connector_c}) or not at all (\protect\subref{fig:memory_ginseng_b}), yet always introduces significant bias in solution estimates. These results suggest that memory is crucial for estimation accuracy, and finite memory does not offer a favorable bias-performance trade-off for either VWoS (\protect\subref{fig:memory_mushroom_a}, \protect\subref{fig:memory_ginseng_b}) or VWoSt (\protect\subref{fig:memory_connector_c}).}
	\label{fig:finite_memory}
\end{figure}

Walk memory $\memory$ is a consequence of the law of total expectation in the derivation of the volumetric BIE \cref{eqn:vbie}. Memory arises likewise in the derivation of the volume rendering equation (\cref{sec:rendering}); however, a memoryless approximation is commonplace in volume rendering algorithms \citep{bitterli2018framework}, especially for exponential media. Motivated by this precedent, we experimented with a memoryless variant of VWoS (i.e., always performing unconditional closest point sampling). We found that this approximation leads to highly inaccurate solution estimates \emph{and} much worse runtime performance: Walks rarely terminate inside the volume while still sampling step sizes that decrease super-exponentially with medium density; hence average walk length increases dramatically. Thus, memory appears to be more important for our algorithms than for rendering ones---we expand on this point in \cref{sec:rendering}. 

To help assess the performance impact of memory, we first visualize in \cref{fig:memory_statistics} statistics relating walk length to the sizes (numbers of balls) $\emptyspacesize_k$ of the empty-ball memory $\emptyspace\paren{\memory_k}$, and $\particlesize_k$ of the  sampled-particle memory $\particles\paren{\memory_k}$, for the experiments in \cref{sec:ensemble_averaging}. $\emptyspacesize_k$ always equals the walk length $k$---each walk step adds a new empty ball. By contrast, the growth of $\particlesize_k$ with walk length depends on the density and type of boundary conditions on the particles: For Dirichlet boundary conditions, each walk step adds at most one---and often zero---new particle during a closest point query (\cref{sec:conditional_sampling}). For Neumann boundary conditions, each walk step may add several particles during a closest silhouette point query (\cref{sec:conditional_silhouette_sampling}). In both cases, higher particle density leads to faster growth of $\particlesize_k$.

Inspired by \citet{seyb2024stochastic}, we also experiment with a variant of VWoS that uses \emph{finite} memory---only storing the most recent empty ball and sampled particle---to mitigate the overhead of conditional sampling as memory grows. \Cref{fig:finite_memory} compares this variant with standard VWoS that uses full memory. Using finite memory introduces significant bias and provides marginal to no performance improvements, suggesting that finite memory of any size does not offer a favorable bias-performance trade-off---larger sizes of finite memory will provide only smaller performance improvements. Performance does not improve more significantly because computational cost is dominated by geometric queries against the medium boundary, which are repeated at each walk step even if step size is very small. The overhead from these repeated queries becomes worse as particle density, and therefore walk length, increases (\cref{fig:memory_statistics}, \emph{top row}). However, if this overhead is reduced---for example, by leveraging the spatial coherence of steps in volumetric walks---then it may become useful to reconsider the use of finite memory. Additionally, finite memory could help improve parallelism on memory-constrained GPU devices, due to improved memory management. We defer further exploration of finite memory to future work. 

\section{Simulation of natural phenomena}\label{sec:physics}

To showcase the ability of our algorithms to deal with a variety of simulation settings with complex microparticle geometry, we use them for simulation of two model natural phenomena. 

\subsection{Electrostatics near biological membranes}\label{sec:membrane}

In \cref{fig:Membrane}, we use VWoS to model electrostatic potentials in systems with complex cellular geometries. Simulating such potentials and their effect on biomolecules is common in biochemistry \citep{davis1990electrostatics,gilson1988calculating}.
However, accurately accounting for the complex molecular geometry involved in these simulations poses significant computational challenges. To manage this complexity, homogenization approaches abstract away the fine details of molecular geometry using simplified models based on asymptotic cases. One such model, the Debye-Hückel framework \citep{andelman1995electrostatic}, is commonly used to describe the screening of electrostatic potentials by ions. Though effective in some scenarios, these models often lead to inaccurate predictions, especially near membrane surfaces where the size of ions relative to the geometry becomes critical \citep{davis1990electrostatics}.

The setup in \cref{fig:Membrane} assumes constant particle density (expected 1.9 million particles per configuration). The concentrations of sodium ($\mathrm{Na}^+$) and chloride ($\mathrm{Cl}^-$) ions depend on the distance to the nearest charged surface: $\mathrm{Cl}^-$ ions are more concentrated near positively charged surfaces, and $\mathrm{Na}^+$ ions are more concentrated near negatively charged surfaces. The boundary conditions ensure neutrality (zero charge) away from the surface, and the charge reaches a maximum value of $\pm \qty{0.25}{\volt}$ on the surface. Compared to running WoS in this setup with the participating medium removed, VWoS increases runtime by only $15\%$. In return for this slight overhead, VWoS produces solutions that greatly improve modeling accuracy of electrostatic potentials due to particle-membrane interactions.

\begin{figure*}[t]
	\centering
	\includegraphics{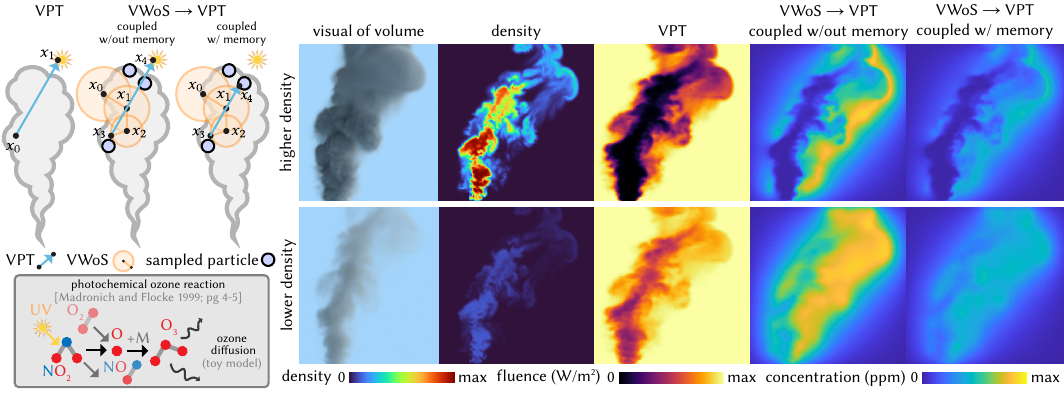}
	\caption{We couple VWoS and VPT to model diffusion and light transport (resp.) in a proof-of-concept atmospheric photochemical system: a cloud \citep{pbrtv3scenes} inside which fluence due to multiply scattered sunlight generates ozone, which then diffuses throughout the cloud (\emph{first and second column}). We estimate ozone concentration (\emph{fourth and fifth column}) by performing VWoS walks that, upon termination, use VPT paths to estimate a Dirichlet boundary condition equal to fluence (\emph{third column}). As both the VWoS walk and VPT path interact with the same microparticle geometry, the memory accumulated from the walk must carry over to the path. Correctly coupling with memory has a non-trivial impact on the estimated ozone concentration (\emph{fifth column}).}
	\label{fig:smoke}
\end{figure*}

This experiment highlights the advantages of VWoS over ensemble averaging, and in particular its ability to simulate complex microparticle geometry without the need to repeatedly sample entire particle configurations. For example, in \cref{fig:Membrane}, computing the electrostatic potential on a $256 \times 256$ slice plane takes $\qty{11}{\sec}$ with ensemble averaging, and only $\qty{1}{\sec}$ with VWoS---a speedup of more than an order of magnitude. The performance improvement is even greater when solution estimates are needed only along a line (e.g., to compute the screening plots in the insets), in which case VWoS is more than $10,000\times$ faster than ensemble averaging.

\subsection{Photochemical effect in clouds}\label{sec:coupled_volumetric}

As we discuss in \cref{sec:pbm,sec:rendering}, the PBM we use to derive VWoS and VWoSt also underlies volume rendering algorithms such as volumetric path tracing (VPT) \citep{novak2018monte}. These algorithms are commonplace in scientific applications that model light transport inside participating media such as clouds or tissue. Combining VWoS and VWoSt with volume rendering creates opportunities for \emph{coupled simulation} of light transport and other physical phenomena (e.g., diffusive effects) in these media.
Such a combination continues recent work on coupled Monte Carlo simulation in deterministic geometry---e.g., combining WoS and path tracing to simulate multimodal heat transfer \citep{Bati:2023:Coupling}.

As an example of such an opportunity, in \cref{fig:smoke}, we couple VWoS and VPT to simulate a simplified atmospheric photochemical system. In such systems, sunlight drives the production of pollutants such as ozone, which then diffuse in the atmosphere \citep{madronich1999role, seinfeld2016atmospheric}. Clouds in particular play a prominent role in atmospheric photochemistry \citep{hall2018cloud,bianco2020photochemistry}, but simulation of the photochemical effect in them is challenging, due to their incredibly complex microparticle geometry (billions of water droplets \citep{gryspeerdt2022impact}). 

Coupled volumetric rendering and PDE simulation can help overcome this challenge. As a proof-of-concept demonstration, we use a cloud model from \textsc{pbrt-v3} \citep{pbrtv3scenes} to set up a photochemical-effect simulation problem. We use VPT to model sunlight (light transport), and VWoS to model ozone concentration as a Dirichlet Laplace BVP \labelcref{eqn:bvp_dirichlet}---an approximation to more accurate models of ozone diffusion \citep{hanna1982handbook}. At each point in the cloud, we set the Dirichlet boundary data equal to the mean incident \emph{fluence} $\influence$ ($\si{\watt\per\meter\squared}$) due to a directional light source modeling the Sun. 

Altogether, we can estimate ozone concentration inside the cloud by running VWoS walks that, upon termination, switch to VPT light paths to estimate incident fluence. Importantly, when tracing these paths, we must account for memory about the medium accumulated during the walk: as light transport and diffusion occur in the same medium, the fluence estimation must be conditioned on the same information as the ozone concentration estimation, i.e., $\meansolution(\conditionals{\point_k}{\memory_k}) = \influence\paren{\conditionals{\point_k}{\memory_k}}$. We handle this conditional evaluation by thinning the medium density with respect to previously sampled empty balls, and occluding rays hitting previously sampled particles. As \cref{fig:smoke} shows, accounting for memory when coupling VWoS and VPT captures shadowing effects that drastically impact the mean solution. Additionally, if we scale the heterogeneous cloud density, the ozone concentration changes in a manner resembling light transport variation in optically dense versus thin media.

\section{Limitations and future work}\label{sec:conclusion}

We introduced the problem of \emph{PDE simulation in participating media} as a framework for computationally modeling natural phenomena involving complex microparticle geometry. We described how to rigorously model such geometry stochastically using the \emph{Poisson Boolean model}, which enabled us to develop sampling procedures that serve as stochastic counterparts of common geometric queries. We additionally developed the \emph{volumetric walk on spheres} (VWoS) and \emph{volumetric walk on stars} (VWoSt) algorithms that use these sampling procedures for Monte Carlo estimation of the solution to this simulation problem. Our algorithms directly generalizes the standard walk on spheres and walk on stars algorithms, maintaining their structure and sharing their attractive features.

Clearly, these contributions are just first steps, aimed to set the groundwork for exploring the interaction between PDEs and participating media. By bridging ideas from rendering, PDE simulation, and stochastic geometry, the problem setting we introduced presents opportunities for future research in several exciting directions.

\paragraph{Sampling algorithms} A core component of our VWoS algorithm is closest point sampling with the thinning procedure of \cref{alg:closest_sampling}. This procedure often becomes the performance bottleneck of our algorithm, for example in volumes with highly concentrated density. A similar bottleneck behavior occurs in volume rendering, where an analogous algorithm, \emph{delta tracking}, is used for free-flight distance sampling \citep{coleman1968mathematical,raab2006unbiased}---we elaborate on this analogy in \cref{sec:rendering}. These performance issues have motivated extensive rendering research on improved sampling algorithms (as well as algorithms for the closely related problem of transmittance estimation \citep{georgiev2019integral,kettunen2021unbiased,kutz2017spectral}), for example by making better use of intermediate sampled points \citep{novak2014residual} or using multiple-importance sampling \citep{miller2019null}. Other approaches address performance issues by using adaptive or progressive majorants \citep{misso23progressive} instead of a global one. All these approaches can potentially be adapted to VWoS, and help bring about similar performance improvements.

\paragraph{Microparticle geometry models} We focused on the simplest form of a Boolean model that uses independent, spherical, and fixed-size particles. However, many phenomena involve participating media with particles that violate these assumptions: they can have non-spherical particles (biological materials such as seeds and pollen, soil and other sediment); particles of wideband size distributions (powders, sand grains); or non-independent particles that form regular structures (crystals), or can repel and attract each other (ions, colloids). Extending our methods to these more varied participating media requires using more general Boolean models for microparticle geometry. Poisson Boolean models with non-spherical particles or particles of random sizes share many of the same properties we used in \cref{sec:pbm} \citep[Chapter 17]{last2017lectures}. Thus, we expect that simulating these models should require only modest modifications to VWoS. Other Boolean models  sample correlated particle centers from non-Poisson point processes \citet[Chapter 5]{chiu2013stochastic} (for example, Matérn \citep{stoyan1985one} or Gibbs \citep{sabatini2024special}). Simulation of such models is more challenging, due to the lack of efficient closest point sampling procedures. Efforts to generalize VWoS to more general Boolean models can benefit from insights from volume rendering, where there already exist variants of VPT for participating media with multi-sized \citep{frisvad2007computing}, anisotropic \citep{jakob2010anisotropy,heitz2015sggx}, or correlated \citep{jarabo2018radiative,bitterli2018framework,d2018reciprocal,d2019reciprocal} particles.

\paragraph{Other stochastic geometry models} Recent work in computer graphics has shown it can be beneficial to model macroscopic object-level geometry stochastically, for example to account for uncertainty in 3D acquisition or facilitate shape optimization \citep{vicini2021non,miller2024oav}. Particularly appealing in this context are stochastic implicit surface representations based on Gaussian processes, thanks to their rich mathematical properties \citep{sheffield2007gaussian} and relationship to mature geometry processing algorithms \citep{sellan2022stochastic,sellan2023neural}. Algorithms for rendering such representations have emerged recently \citep{seyb2024stochastic}, and can help inform research on analogous algorithms for PDE simulation.

\paragraph{More general PDEs} Our volumetric WoS and WoSt algorithms can solve the same types of PDEs as their counterparts for deterministic geometry---we focused on the Laplace equation, but our algorithms are straightforward to apply to Poisson and screened Poisson equations. However, many phenomena suitable for modeling using participating media involve PDEs out of scope for our algorithms, such as linear elasticity \citep{james1999artdefo}, heat conduction and convection \citep{Hahn:2012:Heat}, Stokes flow \citep{du2020functional}, and Navier-Stokes flow \citep{stam1999stable}. Recent work has introduced Monte Carlo simulation algorithms for some of these PDEs \citep{Rioux-Lavoie:2022:MCFluid,sugimoto2024velocity,Bati:2023:Coupling}, often directly extending WoS or WoSt. Using our theory to develop volumetric variants of these algorithms would greatly expand the type of phenomena we can simulate using participating media.


\begin{acks}
This work was supported by National Science Foundation (NSF) award 2212290; National Institute of Food and Agriculture award 2023-67021-39073; a gift from nTopology; NSF Graduate Research Fellowship DGE2140739 and an NVIDIA graduate fellowship for Bailey Miller; a Packard Foundation Fellowship for Keenan Crane; and Alfred P. Sloan Research Fellowship FG202013153 for Ioannis Gkioulekas. Rohan Sawhney thanks Ken Museth for supporting this work. The protein model (\cref{fig:Membrane}) is from user \emph{QuadroFlow} on \textsc{TurboSquid}. The ginseng root model (\nameCrefs{fig:Membrane}~\labelcref{fig:homogenization_comparison}(\subref{fig:ginseng_d}--\subref{fig:ginseng_f}), \labelcref{fig:finite_memory}(\subref{fig:memory_ginseng_b})) is from user \emph{3dror} on \textsc{TurboSquid}. The mushroom model (\nameCrefs{fig:Membrane}~\labelcref{fig:homogenization_comparison}(\subref{fig:mushroom_a}--\subref{fig:mushroom_c}), \labelcref{fig:finite_memory}(\subref{fig:memory_mushroom_a})) and the connector model (\nameCrefs{fig:Membrane}~\labelcref{fig:neumann_validation}, \labelcref{fig:finite_memory}(\subref{fig:memory_connector_c})) are from the \textsc{Thingi10k} dataset \citep{Thingi10K}. The cloud model (\cref{fig:smoke}) is from the \textsc{pbrt-v3} scene repository \citep{pbrtv3scenes}.
\end{acks}

\bibliographystyle{ACM-Reference-Format}
\bibliography{volumetric-wos}

\appendix

\section{Details on the Poisson Boolean model}\label{sec:proof}

We elaborate on the closest point distribution and the related spherical contact distribution in the PBM. We also prove the correctness of conditional closest point sampling using \cref{alg:closest_sampling} with the conditional density of \cref{eqn:conditional_rate}. We use the following property of the Poisson point process \citep[Definition 3.1]{last2017lectures}. 


\begin{prp}[label={pro:ppp}]{Properties of the Poisson point process}{prop}
	We assume that the point set $\centers_\scene \coloneq \curly{\centerpoint_\iterball \in \medium}_{\iterball=1}^\numballs$ is a sample of a \emph{Poisson point process} with rate function $\rate$. Then: 
	\begin{enumerate}[leftmargin=*,label={{P\arabic*}.},ref={{P\arabic*}}]
		\item \label{enu:exponential} For any set $\queryset \subset \medium$,
		\begin{align}
			\Prob\curly{\queryset \cap \centers_\scene = \emptyset} = \exp\paren{- \int_{\queryset} \rate\paren{\pointalt} \ud \pointalt}. \label{eqn:capacity_pbm}
		\end{align}

		\item \label{enu:independent} For any two sets $\queryset, \querysetalt \subset \medium$, if $\queryset \cap \querysetalt = \emptyset$, then the events $\queryset \cap \centers_\scene = \emptyset$ and $\querysetalt \cap \centers_\scene = \emptyset$ are independent.
	\end{enumerate}
\end{prp}
Properties \labelcref{enu:exponential,enu:independent} are analogues of the properties of exponential 
and independent (resp.) increments of the Poisson process on the real line. \labelcref{enu:exponential} also explains the term \emph{exponential media} for volumes with stochastic microparticle geometry following the PBM. 

\paragraph{Spherical contact and closest point distributions.} We consider the random closest point $\closestpointmicro\paren{\point} \in \partial\scene$ and random shortest distance $\largestradiusmicro\paren{\point} \coloneq \norm{\point-\closestpointmicro\paren{\point}}$ between a point $\point \in \medium$ and the boundary $\partial\scene$. The random variables $\largestradiusmicro\paren{\point}$ and $\closestpointmicro\paren{\point}$ follow the \emph{spherical contact distribution} and \emph{closest point distribution} \citep[Section 16.3]{last2017lectures}, whose PDFs we denote $\probradius_{\point}$ and $\probcp_{\point}$ (resp.).

We can derive $\probradius_{\point}$ from $\probdc_{\point}$ (\cref{eqn:pdc}), using the relationship $\largestradiusmicro\paren{\point} = \largestradiuscenter\paren{\point} - \size$ (\cref{fig:centers_points}). Doing so requires conditioning on $\largestradiuscenter\paren{\point} \ge \size$ to ensure a positive value for $\largestradiusmicro\paren{\point}$. Requiring $\largestradiuscenter\paren{\point} \ge \size$ is equivalent to $\point \notin \scene$, or $\ball\paren{\point, \size} \cap \centers_\scene = \emptyset$. Then, from \cref{eqn:pdc,eqn:capacity_pbm} and the definition of conditional probabilities:
\begin{align}
	\probradius_{\point}\paren{\radius} &\coloneq \probdc_{\point}\paren{\conditionals{\radius+\size}{\ball\paren{\point, \size} \cap \centers_\scene = \emptyset}} \\
	&= \exp\paren{-\rateset\paren{\point,\radius+\size} + \rateset\paren{\point,\size}} \int_{\partial\ball\paren{\point,\radius+\size}} \rate\paren{\pointalt} \surfMeasure\paren{\pointalt}.\label{eqn:scd}
\end{align}
We can likewise derive $\probcp_{\point}$ from $\probcc_{\point}$ (\cref{eqn:pcc_cond}), this time using the relationship $\closestpointmicro\paren{\point} = \closestcenter\paren{\point} - \size \dirop\paren{\point,\closestcenter\paren{\point}}$ (\cref{fig:centers_points}). From \cref{eqn:pdc,eqn:capacity_pbm} and conditioning on $\largestradiusmicro\paren{\point} = \radius$,
\begin{equation}\label{eqn:pcp_cond}
	\probcp_\point\paren{\conditional{\pointalt}{\radius}} \coloneq \frac{\rate\paren{\pointalt + \size\ \dirop\paren{\point,\pointalt}}}{\int_{\partial\ball\paren{\point,\radius+\size}} \rate\paren{\pointalt} \surfMeasure\paren{\pointalt}} \frac{\paren{\radius+\size}^2}{\radius^2}.
\end{equation}
In the homogeneous case, \cref{eqn:scd,eqn:pcp_cond} simplify to:
\begin{align}
	\probradius_\point\paren{\radius} &\homoeq \exp\paren{-\nicefrac{4}{3}\pi\paren{\paren{\radius + \size}^3-\size^3}\rate} 4 \pi \paren{\radius + \size}^2 \rate, \label{eqn:scd_homo} \\
	\probcp_\point\paren{\conditional{\pointalt}{\radius}} &\homoeq \frac{1}{4 \pi \radius^2}. \label{eqn:pcp_condo_homo}
\end{align}
Comparing \cref{eqn:scd,eqn:pcp_cond} with \cref{eqn:pdc,eqn:pcc_cond} (resp.), we note that the distributions for $\largestradiusmicro\paren{\point}$ and $\closestpointmicro\paren{\point}$ are closely related to those of $\largestradiuscenter\paren{\point}$ and $\closestcenter\paren{\point}$. However, the simpler expressions for the center-based quantities, and in particular the exponential property of $\paren{\largestradiuscenter\paren{\point}}^3$, greatly simplify sampling a closest center relative to directly sampling a closest point (\cref{sec:pbm}).

\paragraph{Proof of conditional sampling.} The statement $\queryset \cap \scene = \emptyset$ is equivalent to $\queryset^{\oplus\size} \cap \centers_\scene = \emptyset$. Using Properties \labelcref{enu:exponential,enu:independent} and the definition of conditional probabilities, we can update \cref{eqn:pdc} as:
\begin{align}
	&\probdc_{\point}\paren{\conditionals{\radius}{\queryset^{\oplus\size} \cap \centers_\scene = \emptyset}} = \exp\paren{-\rateset\paren{\point,\radius}+\int_{\ball\paren{\point,\radius}\cap\queryset^{\oplus\size}} \rate\paren{\pointalt} \ud \pointalt} \nonumber \\
	&\qquad\ \ \ \cdot \paren{\int_{\partial\ball\paren{\point,\radius}} \rate\paren{\pointalt} \surfMeasure\paren{\pointalt}-\int_{\partial(\ball\paren{\point,\radius}\cap\queryset^{\oplus\size})} \rate\paren{\pointalt} \surfMeasure\paren{\pointalt}},\label{eqn:pdc_cond}
\end{align}
which equals the unconditional $\probdc_{\point}$ of \cref{eqn:pdc} computed using the conditional density  $\rate(\conditionals{\cdot}{\queryset})$ in \cref{eqn:conditional_rate}.

\section{Volume rendering}\label{sec:rendering}

It is instructive to compare and contrast VWoS with volume rendering algorithms for exponential media, and in particular its closest analogue, \emph{volumetric path tracing} (VPT). We first overview VPT, then discuss analogies with VWoS. Our discussion of VPT and volume rendering is brief, and we refer to \citet{novak2018monte} and \citet[Chapter 15]{pharr2023physically} for more detailed treatments.

\paragraph{VPT overview.} In rendering, the domain $\domain$ plays the role of the scene. When the scene geometry is deterministic, path tracing computes the incident radiance $\inradiance$ at a point $\point$ and direction $\direction$ recursively, using the conservation law for radiance along a ray:
\begin{equation}\label{eqn:radiance_conservation}
	\inradiance\paren{\point, \direction} = \outradiance(\closestpoint\paren{\point,\direction}, -\direction),
\end{equation}
where $\outradiance$ is outgoing radiance. The point $\closestpoint\paren{\point,\direction}$ is the first intersection with the scene boundary $\partial\domain$ of a ray with origin $\point$ and direction $\direction$. We term the distance $\casting\paren{\point,\direction} \coloneq \Vert\point - \closestpoint\paren{\point,\direction}\Vert$ the \emph{ray distance} along the ray with origin $\point$ and direction $\direction$. Noting that $\closestpoint\paren{\point,\direction} = \point + \casting\paren{\point,\direction} \direction$, \cref{eqn:radiance_conservation} becomes:
\begin{equation}\label{eqn:radiance_conservation_ff}
	\inradiance\paren{\point, \direction} = \outradiance(\point + \casting\paren{\point,\direction} \direction, -\direction).
\end{equation}
When the scene geometry is deterministic, we can compute $\closestpoint\paren{\point,\direction}$ and $\casting\paren{\point,\direction}$ by performing a \emph{ray casting query}.

In a participating medium, the scene includes both deterministic and stochastic geometry as in \cref{eqn:domain}. Volumetric path tracing then computes the incident \emph{mean radiance} $\inmeanradiance$, defined analogously to the mean solution $\meansolution$ in \cref{eqn:mean_solution}.%
\footnote{Notably, the rendering literature typically does not distinguish between mean radiance and radiance, treating the two quantities as interchangeable.} 
Taking the expectation of \cref{eqn:radiance_conservation_ff} and applying the law of total expectation gives us:
\begin{equation}\label{eqn:mean_radiance_conservation_rd}
	\inmeanradiance\paren{\point, \direction} = \int_0^\infty \pdf^\distance(\distance) \outmeanradiance(\conditionals{\point + \distance \direction, -\direction}{\casting\paren{\point,\direction} = \distance}) \ud \distance.
\end{equation}
Here, we used the fact that in a participating medium the ray distance $\casting\paren{\point,\direction}$ is a continuous random variable, indicating its PDF as $\pdf^\distance$. From \cref{eqn:domain}, it follows that:
\begin{equation}\label{eqn:casting_min}
	\casting\paren{\point,\direction} = \min\{\castingmicro\paren{\point,\direction},\castingvol\paren{\point,\direction}\},
\end{equation}
where:
\begin{enumerate*}
	\item $\castingmicro\paren{\point,\direction}$ is the random ray distance to the boundary $\partial\scene$ of the microparticle geometry;
	\item $\castingmicro\paren{\point,\direction}$ is the deterministic ray distance to the boundary $\partial\medium$ of the volume.
\end{enumerate*}

The random ray distance $\castingmicro\paren{\point,\direction}$ follows the \emph{linear contact distribution} \citep[Section 16.3]{last2017lectures} with associated PDF $\probff_{\point,\direction}$ and tail distribution function $\transmittanceff_{\point,\direction}$. The rendering literature uses the terms \emph{free-flight distance} for $\castingmicro\paren{\point,\direction}$, \emph{free-flight distribution} for $\probff_{\point,\direction}$, and \emph{transmittance} for $\transmittanceff_{\point,\direction}$ \citep{bitterli2018framework}. From \cref{eqn:casting_min}, it follows that we can rewrite \cref{eqn:mean_radiance_conservation_rd} as:
\begin{align}\label{eqn:volume_rendering}
	\inmeanradiance\paren{\point, \direction} &= \int_0^{\castingvol}\probff_{\point,\direction}(\distance) \outmeanradiance(\conditionals{\point + \distance \direction, -\direction}{\casting = \distance}) \ud \distance \nonumber \\
	&+ \transmittanceff_{\point,\direction}(\castingvol) \outmeanradiance(\conditionals{\point + \castingvol \direction, -\direction}{\casting = \castingvol}),
\end{align}
where we used the shorthands $\casting$, $\castingmicro$, and $\castingvol$ to simplify notation. \Cref{eqn:volume_rendering} is the \emph{volume rendering equation} (VRE). To estimate $\inmeanradiance$, VPT first samples a free-flight distance $\distance \sim \probff_{\point,\direction}$ and then:
\begin{enumerate}
	\item estimates $\outmeanradiance(\conditionals{\point + \distance \direction, -\direction}{\casting = \distance})$ if $\casting < \castingvol$, effectively applying single-sample Monte Carlo to the integral term;
	\item estimates $\outmeanradiance(\conditionals{\point + \castingvol \direction, -\direction}{\castingvol = \distance})$ if $\casting \ge \castingvol$.
\end{enumerate}
Estimation then proceeds recursively, albeit \emph{without memory}: VPT approximates $\outmeanradiance(\conditionals{\point + \distance \direction, -\direction}{\casting = \distance}) \approx \outmeanradiance(\point + \distance \direction, -\direction)$---a so-called \emph{renewal assumption} \citep{seyb2024stochastic}---and continues to iterate \cref{eqn:volume_rendering} (after first using the in-scattering equation to convert from outgoing to incident radiance).

To make free-flight distance sampling tractable, classical VPT algorithms assume \emph{exponential media}, or equivalently, that the microparticle geometry follows the PBM $\PBM\paren{\rate,\size}$. Using the PBM properties in \cref{pro:ppp} (analogously to the derivation of $\probradius_{\point}$ in \cref{eqn:scd}), as well as an assumption that particle size $\size$ is appropriately small, the transmittance and free-flight distribution become \citep[Section 16.3]{last2017lectures}:
\begin{align}
	\transmittanceff_{\point,\direction}\paren{\distance} &= \exp(-\int_0^\distance 4 \pi \size^2 \rate(\point + \distancealt \direction) \ud \distancealt),\label{eqn:transmittanceff} \\
	\probff_{\point,\direction}\paren{\distance} &= \exp(-\int_0^\distance 4 \pi \size^2 \rate(\point + \distancealt \direction) \ud \distancealt) 4 \pi \size^2 \rate(\point + \distance\direction),\label{eqn:pff}
\end{align}
or in the homogeneous case, analogously to \cref{eqn:scd_homo}:
\begin{align}
	\transmittanceff_{\point,\direction}\paren{\distance} &\homoeq \exp\paren{-4 \pi \size^2 \rate \distance},\label{eqn:transmittanceff_homo} \\
	\probff_{\point,\direction}\paren{\distance} &= \exp\paren{-4 \pi \size^2 \rate \distance} 4 \pi \size^2 \rate.\label{eqn:pff_homo}
\end{align}
We note that the closest analogue of the free-flight distribution $\probff_{\point,\direction}$ in VWoS is the spherical contact distribution $\probradius_{\point}$ (\cref{eqn:scd}). However, under the small $\size$ assumption, $\probff_{\point,\direction}$ is more similar to the shortest distance-to-center distribution $\probdc_{\point}$ (\cref{eqn:pdc}).

It follows that, in the homogeneous case, sampling the free-flight distance requires simply sampling an exponential random variate with rate $4 \pi \size^2 \rate \distance$. In the heterogeneous case, sampling can be done by thinning \citep{lewis1979simulation}, also known as \emph{delta tracking} in the rendering literature: We progressively increase the free-flight distance by increments sampled assuming homogeneous density $\majrate \ge \rate$, until we accept a value with probability $\nicefrac{\rate}{\majrate}$.

\paragraph{Comparison with VWoS} The above discussion helps highlight the many analogies between our VWoS algorithm for PDE simulation in participating media, and VPT for rendering in participating media.
\begin{itemize}[leftmargin=*]
	\item Both algorithms are derived by first transforming a recursive equation through the law of total expectation, then applying single-sample Monte Carlo: the BIE \labelcref{eqn:bie} becomes the BIE in participating media \labelcref{eqn:vbie} for VWoS; and the radiance conservation law \labelcref{eqn:radiance_conservation_ff} becomes the VRE \labelcref{eqn:volume_rendering} for VPT.
	\item Both algorithms replace deterministic geometric queries with sampling operations: VWoS replaces closest point queries with closest point sampling, and VPT replaces ray-casting queries with free-flight distance sampling.
	\item These sampling operations require characterizing associated PDFs: the shortest distace-to-center $\probdc_{\point}$ in VWoS, and the free-flight distribution (or linear contact distribution) $\probff_{\point,\direction}$ in VPT.
	\item Both algorithms use the PBM to make these distributions easy to sample: the cubed shortest distance in VWoS and the free-flight distance in VPT become exponential random variables.
	\item Both algorithms use thinning for sampling under heterogeneous densities: WoS uses \cref{alg:closest_sampling}, and VPT uses delta tracking.
\end{itemize}
\begin{table*}[t]
	\centering
	\caption{\label{tab:scene_parameters} Scene parameters for experiments in \cref{sec:evaluation,sec:physics}. We report the maximum density and corresponding mean free ball radius (average shortest distance to particle centers) of the participating media. We also report the maximum extent (maximum length across all dimensions) of the scenes.}
	\begin{tabular*}{\linewidth}{@{\extracolsep{\fill}} {l}*{5}{c}@{}}
		\toprule
		scene & $\varepsilon$-shell width & particle size $\size$ & maximum density $\rate$ & mean free ball radius & maximum scene extent\\
		\midrule
		\cref{fig:Membrane}, membrane                                            & $1 \times 10^{-4}$    & $1 \times 10^{-3}$&   $1 \times 10^{5}$   & $1.2 \times 10^{-2}$    & $4.0 \times 10^{0}$ \\
		\cref{fig:homogenization_comparison}(\subref{fig:mushroom_a}), mushroom  & $1 \times 10^{-4}$    & $5 \times 10^{-2}$ &  $1 \times 10^2$     & $1.2 \times 10^{-1}$    & $2.3 \times 10^{0}$ \\
		\cref{fig:homogenization_comparison}(\subref{fig:mushroom_b}), mushroom   & $1 \times 10^{-4}$    & $1 \times 10^{-2}$ &  $5 \times 10^2$     & $7.0 \times 10^{-2}$    & $2.3 \times 10^{0}$ \\
		\cref{fig:homogenization_comparison}(\subref{fig:mushroom_c}), mushroom  & $1 \times 10^{-4}$    & $1 \times 10^{-3}$ &  $5 \times 10^3$     & $3.2 \times 10^{-2}$    & $2.3 \times 10^{0}$ \\
		\cref{fig:homogenization_comparison}(\subref{fig:ginseng_d}), ginseng & $1 \times 10^{-4}$    & $2 \times 10^{-3}$ &  $1 \times 10^6$     & $5.5 \times 10^{-3}$     & $4.0 \times 10^{-1}$ \\
		\cref{fig:homogenization_comparison}(\subref{fig:ginseng_e}), ginseng & $1 \times 10^{-4}$    & $2 \times 10^{-3}$ &  $1 \times 10^6$     & $5.5 \times 10^{-3}$    & $2.0 \times 10^{-1}$ \\
		\cref{fig:homogenization_comparison}(\subref{fig:ginseng_f}), ginseng  & $1 \times 10^{-4}$    & $2 \times 10^{-3}$ &  $1 \times 10^6$     & $5.5 \times 10^{-3}$    & $1.0 \times 10^{-1}$\\
		\cref{fig:neumann_validation}(\subref{fig:more_conc_a}), connector & $1 \times 10^{-3}$    & $2 \times 10^{-2}$ &  $5.5 \times 10^5$   & $6.8 \times 10^{-3}$    & $1.8 \times 10^{0}$ \\
		\cref{fig:neumann_validation}(\subref{fig:less_conc_b}), connector & $1 \times 10^{-3}$    & $2 \times 10^{-2}$ &  $2.5 \times 10^7$   & $1.9 \times 10^{-3}$    & $1.8 \times 10^{0}$\\
		\cref{fig:smoke}, cloud & $1 \times 10^{-4}$    & $5 \times 10^{-3}$ &  $5 \times 10^4$     & $1.5 \times 10^{-2}$    & $2.0 \times 10^{0}$\\
		\bottomrule
	\end{tabular*}
\end{table*}
\begin{table*}[t]
	\centering
	\caption{\label{tab:scene_boundary_conditions} Dirichlet boundary data $\boundary$ on the medium boundary $\partial\medium$ and particle boundary $\partial\scene$, for experiments in \cref{sec:evaluation,sec:physics}. $\point_i$ is the $i$-th coordinate of the point $\point \in \R^3$. $\closestpointvol\paren{\point}$ and $\largestradiusvol\paren{\point}$ are the closest point and shortest distance (resp.) between $\point$ and $\partial\medium$.}
	\begin{tabular*}{\linewidth}{@{\extracolsep{\fill}} {l}*{2}{c} @{}}
		\toprule
		scene & medium boundary & particle boundary\\
		\midrule
		\cref{fig:Membrane}, membrane & $\mathrm{texture}\paren{\point}$ & $-0.25 \mathrm{texture}(\closestpointvol\paren{\point}) \exp(-200 \largestradiusvol\paren{\point}^2)$\\
		\cref{fig:homogenization_comparison}, mushroom & $\mathrm{texture}\paren{\point}$ & $0$\\
		\cref{fig:homogenization_comparison}, ginseng & { $0.5 (\cos(2 \exp(-2 (\point_2 - 1.75)) \point_0) \cos(2 \exp(-2 (\point_2 - 1.75)) \point_1) - 1.75)$} & $0$\\
		\cref{fig:neumann_validation}, connector & { \textbf{if} $\point_0 < 0$ \textbf{then} $0.5 \cos(10 \point_2) + 0.5$ \textbf{else} $0.5 \cos(10 \point_2) - 0.5$} & $0$\\
		\cref{fig:smoke}, cloud & 0 & fluence $\influence\paren{\point}$\\
		\bottomrule
	\end{tabular*}
\end{table*}
At the same time, there are important differences.
\begin{itemize}[leftmargin=*]
	\item Perhaps the most salient difference relates to \emph{memory}: VWoS has full memory by conditioning, during recursion, on the outcomes of sampling operations at previous steps. VPT has no memory, ``forgetting'' those previous outcomes. Ignoring memory is a reasonable approximation in VPT because linear segments along a light path are unlikely to overlap, making it unnecessary for a segment to condition on the empty space carved by a previous one. By contrast in VWoS, balls along a random walk are very likely to overlap, requiring memory for accurate estimation (\cref{sec:memory_ablation}).
	\item A more subtle difference becomes evident when we compare the BIE in participating media \labelcref{eqn:vbie} with the VRE \labelcref{eqn:volume_rendering}: The former conditions on closest \emph{points}, whereas the latter conditions on shortest ray \emph{distances}. This difference stems from the fact that whereas conditioning on the shortest ray distance fixes a unique closest intersection point, conditioning on the shortest distance only fixes the closest point to lie on a sphere, requiring additional area sampling on that sphere (\cref{pro:polar} and \cref{alg:closest_sampling}). 
\end{itemize}
Reconciling these differences creates future research opportunities, for example towards volume rendering algorithms that use memory for improved estimation accuracy, or towards VWoS variants that use only shortest distance sampling for improved efficiency.

Lastly, we note that volume rendering applications typically specify participating media through their \emph{extinction coefficient} $\sigma_t\paren{\point} \coloneq 4 \pi \size^2 \rate\paren{\point}$ that appears in \crefrange{eqn:transmittanceff}{eqn:pff_homo}, rather than their density $\rate\paren{\point}$. The closed-form relationship between the two allows us to reuse abundant publicly available volume models for rendering also in PDE simulation, as we do in \cref{sec:coupled_volumetric}.

\section{Scene parameters}\label{sec:parameters}

We report scene parameters (\cref{tab:scene_parameters}) and boundary conditions (\cref{tab:scene_boundary_conditions}) for all experiments in \cref{sec:evaluation,sec:physics}.

\end{document}